\documentclass[aps,twocolumn,superscriptaddress,floatfix,amsmath,amssymb]{revtex4}

\usepackage[normalem]{ulem}
\usepackage[usenames,dvipsnames]{color}
\usepackage[nointegrals]{wasysym}
\usepackage{siunitx}
\usepackage{graphicx}
\usepackage{bbm}
\usepackage{ifthen}

\definecolor{gray}{rgb}{.6,.6,.6}
\definecolor{darkyellow}{rgb}{.6,.5,0}
\definecolor{darkgreen}{rgb}{0,.6,0}

\DeclareRobustCommand{\void}[1]{}

\newcommand{\affiliationMunich}{\affiliation{Center for NanoScience \& Fakult\"at f\"ur Physik, LMU-Munich, 80539 M\"unchen, Germany}}
\newcommand{\affiliationRegensburg}{\affiliation{Fakult\"at f\"ur Physik,  Universit\"{a}t Regensburg, 93040 Regensburg, Germany}}
\newcommand{\affiliationAugsburg}{\affiliation{Institut f\"ur Physik, Universit\"at Augsburg, 86135 Augsburg, Germany}}
\newcommand{\affiliationETH}{\affiliation{Solid State Physics Laboratory, ETH Zurich, 8093 Zurich, Switzerland}}
\newcommand{\affiliationMadrid}{\affiliation{Instituto de Ciencia de Materiales de Madrid, CSIC, 28049 Madrid, Spain}}

\newcommand{\ud}{ \uparrow \downarrow }

\newcommand{\up}{{\uparrow}}
\newcommand{\down}{{\downarrow}}

\newcommand{\abs}[1]{\left\vert #1 \right\vert} 

\newcommand{\fig}[2]{Fig.~\ref{#1}\uppercase{#2}}
\newcommand{\twofigs}[3]{Figs.\ \ref{#1}\uppercase{#2} and \uppercase{#3}}

\newcommand{\figs}[3]{Figs.\ \ref{#1}\uppercase{#2}--\uppercase{#3}}
\newcommand{\sect}[1]{Sec.~\ref{#1}}

\newcommand{\suppl}{supplementary material}

\newcommand{\app}[1]{Appendix \ref{#1}}

\newcommand{\Tlr}{\ensuremath{\text{T}_{11}}}

\newcommand{\Slr}{\ensuremath{\text{S}_{11}}}
\newcommand{\Sll}{\ensuremath{\text{S}_{20}}}
\newcommand{\Tll}{\ensuremath{\text{T}_{20}}}

\newcommand{\mB}{\ensuremath{\mu_\text{B}}}

\newcommand{\kT}{\ensuremath{k_\text{B}T}}

\begin{document}


\title{Characterization of Qubit Dephasing by Landau-Zener Interferometry}

\author{F. Forster}\thanks{These authors contributed equally to this work.}\affiliationMunich
\author{G. Petersen\footnotemark[1]}\affiliationMunich
\author{S. Manus}\affiliationMunich
\author{P. H\"anggi}\affiliationAugsburg
\author{D. Schuh}\affiliationRegensburg
\author{W. Wegscheider}\affiliationRegensburg\affiliationETH
\author{S. Kohler}\affiliationMadrid
\author{S. Ludwig}\affiliationMunich
\date{\today}

\pacs{Valid PACS appear here}
\begin{abstract}
Controlling coherent interaction at avoided crossings is at the heart of
quantum information processing. The regime between sudden switches and adiabatic transitions is
characterized by quantum superpositions that enable interference experiments.
Here, we implement periodic passages at intermediate speed in a GaAs-based
two-electron charge qubit and observe Landau-Zener-St\"uckelberg-Majorana (LZSM)
quantum interference of the resulting superposition state. We demonstrate that
LZSM interferometry is a viable and very general tool
to not only study qubit properties but beyond to decipher decoherence caused
by complex environmental influences. Our scheme is based on straightforward
steady state experiments. 
The coherence time of our two-electron charge qubit is limited by
electron-phonon interaction. It is much longer than previously reported for
similar structures.
\end{abstract}

\maketitle

\begin{figure*}
\begin{center}
\includegraphics[width=.9\textwidth]{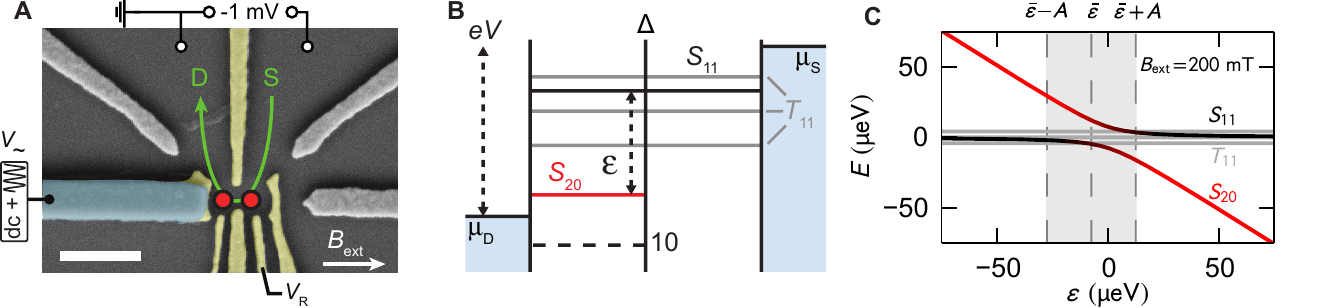}
\caption{\textbf{Experimental setup.}
(\textbf{A}) Scanning electron micrograph showing Ti/Au gates on the surface of
a GaAs\,/\,AlGaAs heterostructure ($500\,$nm scale bar). Six of the Ti/Au gates
(yellow) are used to laterally define a DQD in the 2DES 85\,nm beneath the
surface, the others are grounded. A cobalt single-domain nanomagnet (blue) produces an inhomogeneous
magnetic field which slightly mixes singlet and triplet states of the DQD.
(\textbf{B}) Typical situation in our two-electron DQD: vertical lines indicate
tunable tunnel barriers, horizontal lines chemical potentials, blue areas the
degenerate 2DES leads. The voltage $V = (\mu_\text{S} - \mu_\text{D})/e$
causes a single-electron tunneling current (green arrow in panel A).
(\textbf{C}) Energy diagram of the relevant two-electron DQD eigenstates.
Singlets (the qubit states) are represented as black and red lines; triplets,
which are Zeeman split, as gray lines.  Rf-modulation of the gate voltage
$V_\sim$ (panel A) results in a modulated detuning $\epsilon(t)$, indicated by
gray shading.
\label{figure1}
}
\end{center}
\end{figure*}

LZSM interferometry is a double-slit kind experiment which, in principle, can be
realized with any qubit, while the specific measurement protocol might vary. Our
system is a charge qubit based on two-electron states in a lateral double
quantum dot (DQD) embedded in a two-dimensional electron system (2DES)
(\fig{figure1}{}).
Source and drain leads at chemical potentials $\mu_\text{S,D}$, each tunnel
coupled to one dot, allow
current flow by single-electron tunneling. Applying the voltage
$V=(\mu_\text S-\mu_\text D)/e=1\,$mV across the DQD (\fig{figure1}{b}) we
use this current to detect the steady-state properties of the driven
system.  We interprete the singlets, $\Slr$ (one electron in
each dot) and $\Sll$ (two electrons in the left dot), as qubit states. They
form an avoided crossing (\fig{figure1}{c}), described by the Hamiltonian
\begin{equation}\label{hamiltonian}
 H_\text{qubit}=
\begin{pmatrix}
0&\Delta/2\\
\Delta/2&-\epsilon(t)
\end{pmatrix}\,,
\end{equation}
where we consider a variable energy detuning $\epsilon(t)$ and a constant
inter-dot tunnel coupling tuned to $\Delta\simeq13\,\mu$eV, corresponding to a
clock speed of $\Delta/h\simeq3.1\,$GHz, where $h$ is the Planck constant.

Let us first discuss a single sweep through the avoided crossing at
$\epsilon=0$: as shown back in 1932 independently by Landau, Zener, Stückelberg,
and Majorana it brings the qubit into a superposition state
\cite{Landau1932a,Zener1932a,Stueckelberg1932a,Majorana1932a}, the electronic
analog to the optical beam splitter.  The probability to remain in the initial
qubit state, $P_\text{LZ} = \exp(-\pi\Delta^2/2\hbar v)$, thereby grows with the
velocity $v=\text d \epsilon/\text d t$, here assumed to be constant
\cite{Landau1932a,Zener1932a,Stueckelberg1932a,Majorana1932a}.
Because the relative phase between the split wavepackets depends on their energy evolutions, repeated passages by a periodic modulation
$\epsilon(t)=\bar\epsilon+A\cos(\Omega t)$, give rise to so-called LZSM quantum
interference \cite{Zener1932a, Stueckelberg1932a, Majorana1932a,Oliver2005a,Sillanpaa2006a,Wilson2007b,Berns2008a,Shevchenko2010a,Huang2011,Stehlik2012a,Dupont-Ferrier2013,Cao2013,Nalbach2013,Ribeiro2013}. We present a
breakthrough which makes LZSM interferometry a powerful tool: it is based on
systematic measurements together with a realistic model, which explicitly
includes the noisy environment. We demonstrate how to decipher the detailed
qubit dynamics and directly determine its decoherence time $T_2$ based on
straightforward steady state measurements.


Keeping the experiment simple we detect the dc-current $I$ through the DQD. It
involves electron tunneling giving rise to the configuration cycle
$(1,0)\to(1,1)\leftrightarrow(2,0)\to(1,0)$, where pairs of digits refer to the
number of electrons charging the (left, right) dot (\fig{figure1}{b}).  The
energetically accessible two-electron states include the singlets $\Slr$ and
$\Sll$ but also three triplets $\Tlr$ (\twofigs{figure1}{b}{c}).  These
triplets are likely occupied during $(1,0)\to(1,1)$ and their
decay via a spin-flip, $\Tlr$ $\to$  $\Slr$, is hindered by Pauli-spin
blockade \cite{Ciorga2000a,Ono2002}. This suppresses the transition
$(1,1)\to(2,0)$ and thereby limits the current.
To nevertheless quickly initialize the qubit and generate a measurable current
we lift the blockade using an on-chip nanomagnet (\fig{figure1}a)
\cite{Petersen2013a}.  $I$ is proportional to the occupation probability of
$\Sll$ and serves as destructive qubit detector.

As it is possible to tune the relative couplings and the mean detuning
$\bar\epsilon$ of the singlet-singlet and singlet-triplet crossings by gate
voltages and magnetic fields, our two-electron DQD opens two interesting
perspectives: (i) LZSM interferometry involving multiple avoided crossings and
(ii) coherent Landau-Zener transitions between our charge qubit and the recently
very successful spin-based qubits \cite{Bluhm2011}.

\begin{figure*}
\begin{center}
\includegraphics[width=.85\textwidth]{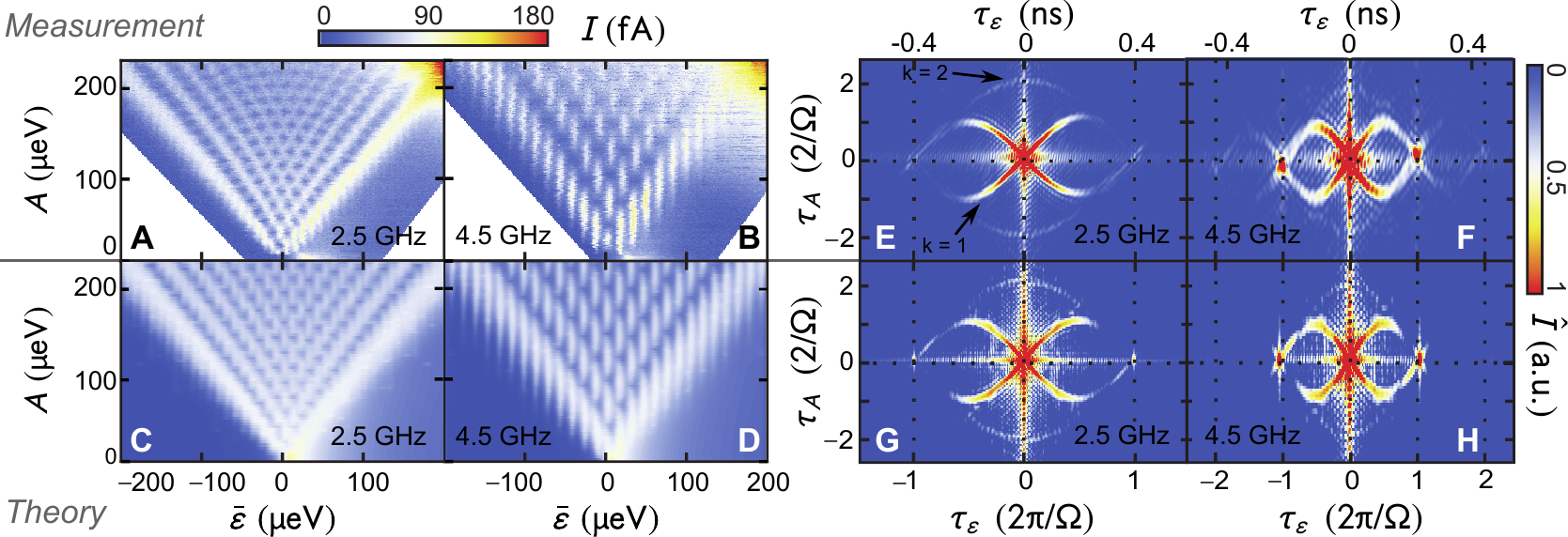}
\caption{\textbf{LZSM interference.}
(\textbf{A} and \textbf{B}) Measured current through
the DQD as a function of mean detuning $\bar\epsilon$ and modulation amplitude
$A$ for two modulation frequencies at $T\simeq20\,$mK. 
(\textbf{C} and \textbf{D}) Same as (A) and (B) but numerically calculated for
realistic conditions.  
(\textbf{E}--\textbf{H}) Two-dimensional numerical Fourier transformed
($A\to\tau_A$, $\bar\epsilon\to\tau_\epsilon$, $I\to\widehat I$) of measurements
(upper panels) and theory (lower panels). The shape of the sinusoidal branches
of enhanced $\widehat I$ is determined by $\Omega$, see Eq.~\eqref{eq:lemons}. Their decay with increasing
$\tau_\epsilon$ encodes dephasing and decoherence.  The horizontal and
vertical lines of enhanced amplitude at $\tau_A=0$ and $\tau_\epsilon=0$ are
artefacts caused by the finite region of data being transformed.
\label{figure2}
}
\end{center}
\end{figure*}

Concentrating on the two-electron charge qubit, in \twofigs{figure2}{a}{b}
we display LZSM interference patterns measured at
$T_\text{2DES}\simeq20\,\text{mK}$ for two different modulation frequencies $\Omega/2\pi$.
Within the triangle defined by $A\gtrsim|\bar\epsilon|$, the qubit is
periodically driven through the avoided crossing and the current oscillates
between zero and distinct maxima indicating destructive and constructive
interference \cite{Kayanuma1994a,Shevchenko2010a}.  An interpretation based on
photon-assisted tunneling (PAT), which is for $\hbar\Omega\gtrsim\Delta$ fully
equivalent to the LZSM picture discussed above, facilitates quantitative
predictions: using Floquet scattering theory \cite{Strass2005b} we find
\begin{equation}
\label{LZSbasic}
I(\bar\epsilon,A) =
\frac{e}{\hbar}\frac{\Gamma_\text{in}\Gamma_\text{out}}{4\gamma}
\sum_{n=-\infty}^\infty
\frac{\Delta_n^2}{(\bar\epsilon-n\hbar\Omega)^2+\Delta_n^2+\gamma^2}\,,
\end{equation}
per spin projection,
where $\Gamma_\text{in}$ is the qubit initialization rate $(1,0)\to \Slr$ and
$\Gamma_\text{out}$ the decay rate $(2,0)\to(1,0)$ and  $\gamma =
\frac{1}{2}(\Gamma_\text{in}+\Gamma_\text{out})$. The interdot tunnel coupling
is renormalized with the $n$th-order Bessel function $J_n$ of the first kind:
$\Delta_n = J_n(A/\hbar\Omega)\Delta$. Eq.~\eqref{LZSbasic} predicts Lorentz-shaped current maxima of width
$\delta\bar\epsilon=\sqrt{\Delta_n^2+\gamma^2}$ at the $n$-photon resonances
$\bar\epsilon = n\hbar\Omega$, modulated by $J_n^2(A/\hbar\Omega)$ as function
of $A$.  This scattering approach provides an appealing physical picture and
describes the main features of the measured LZSM patterns as can be easily seen
for the high frequency limit $\hbar\Omega\gg\delta \bar\epsilon$ (\suppl: Fig.\
\ref{fig:pat}). For lower $\Omega$, the distance between current peaks is smaller
and, hence, the broadened resonances tend to merge (\fig{figure2}{a}).  

The visibility of the LZSM pattern (i) depends on frequency and amplitude via
the Landau-Zener probability $P_\text{LZ}$ (captured in Eq.~\eqref{LZSbasic} by
$\Delta_n$), is (ii) strongest for $\Gamma_\text{in}\simeq\Gamma_\text{out}$ and
is (iii) diminished for $\Delta<\gamma$, where the qubit decay is faster than
its clock-speed.  However, Eq.~\eqref{LZSbasic} fails to predict the qubit coherence
time as it ignores environmental noise. The nevertheless qualitative consent
indicates that environmental noise can be treated perturbatively. In this
spirit, we developed a complete model which goes beyond Eq.~\eqref{LZSbasic} by
explicitly including all energetically accessible states of our driven DQD and,
importantly, decoherence within a system-bath approach.

An evident source of decoherence is the interaction of the qubit electrons with
bulk phonons \cite{Granger2012a} which entails quantum fluctuations to the DQD level energies. It
enters our theory as dissipation kernel with a dimensionless electron-phonon
coupling strength $\alpha_Z$ (\suppl: \ref{suppl:system-lead-bath}, \ref{suppl:charge_qubit}) derived from a system-bath approach
becoming the spin-boson model in the qubit subspace
\cite{Hanggi1990a,Makhlin2001b}.  We assume for the coupling an Ohmic spectral density which is justified by geometry considerations (\suppl: \ref{suppl:system-bath_hamiltonian}) and also a
posteriori by a surprisingly good agreement with our experimental results. 

The second environmental component of our model is charge noise, 
well known to
cause low frequency fluctuations of the local confinement potential in
semiconductor heterostructures
\cite{Fujisawa2000,Pioro-Ladriere2005,Buizert2008,Yacoby2013}.  Being slow
compared to all relevant time scales of our experiment, they can be treated as
static disorder leading in the ensemble average to
an inhomogeneous, Gaussian broadening of width $\lambda^\star$. 

To determine the key parameters $\lambda^\star$ and $\alpha_Z$ we solve the
Bloch-Redfield master equation self-consistently using Floquet theory and numerically model our data (\suppl: \ref{Bloch-Redfield-Floquet_theory}) 
The optimized result is displayed in
\twofigs{figure2}{C}{D} with $\lambda^\star=3.5\,\mu$eV and
$\alpha_Z=1.5\cdot10^{-4}$.  Below, we illustrate the self-consistent procedure
by first determining $\lambda^\star$ based on the final value of $\alpha_Z$ and
then evaluating $\alpha_Z$ using the final value of $\lambda^\star$.

\begin{figure*}
\begin{center}
\includegraphics[width=.90\textwidth]{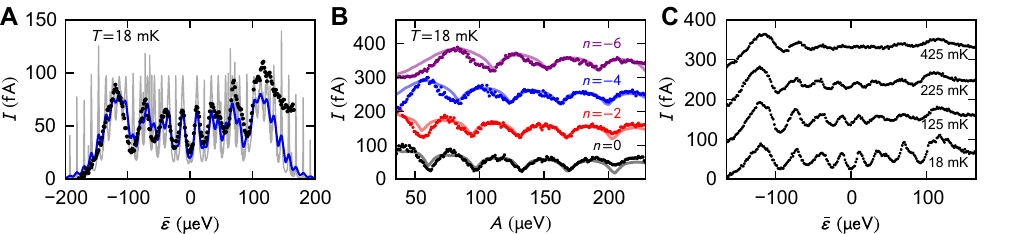}
\caption{\textbf{Raw data analysis.}
Dots are measured at $\Omega/2\pi=2.75\,$GHz, lines numerical data for $\alpha_Z=1.5\cdot10^{-4}$ and $\lambda^\star=3.5\,\mu$eV, only the gray line in panel A is for $\lambda^\star=0$.
(\textbf A) Horizontal slice through a LZSM pattern: $I(\bar\epsilon)$ for a constant
$A=130\,\mu$eV. 
(\textbf B) Vertical slices through a LZSM pattern: $I(A)$ for $\bar\epsilon/
(\hbar\Omega)=0$, $-2$, $-4$, $-6$.
 (\textbf C) Measured data as in panel A for various temperatures. 
}
\label{figure3}
\includegraphics[width=.90\textwidth]{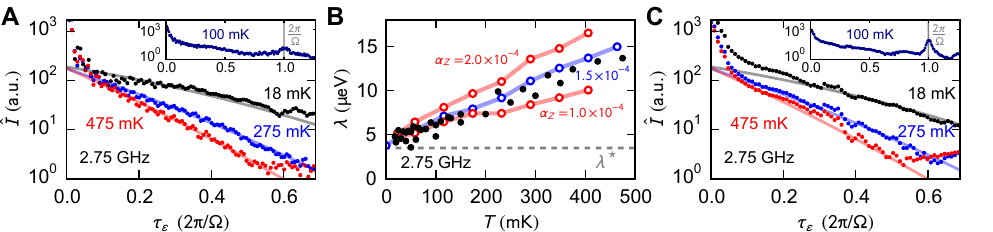}
\caption{\textbf{Electron-phonon coupling.}
\textbf{(A)} Decaying region of the measured $\widehat
I(\tau_\epsilon,\tau_A)\big|_\text{lemon}$ for three temperatures (dots). Lines
are generated using Eq.~\eqref{eq:Fourier} for $\lambda^\star=3.5\,\mu$eV and
 $\lambda$ as a fit parameter. The inset shows a broader region including
maxima at $\tau_\epsilon=0$,  $2\pi/\Omega$.
\textbf{(B)} Measured decay rate $\lambda(T)$ (black) and corresponding
numerical data (colored) based on $\lambda^\star=3.5\,\mu$eV (indicated as horizontal line) and $\alpha_Z=1.0$, $1.5$, $2.0\times10^{-4}$.
\textbf{(C)}
Analog to panel A but based on numerical calculations. Solid lines are identical to those in A. The numerical resolution is based on 100 data points sampling the Gaussian broadening in $\bar\epsilon$ of width $\lambda^\star=3.5\,\mu$eV.
\label{figure4}
}
\end{center}
\end{figure*}

\fig{figure3}{A}
displays $I(\bar\epsilon)$ for $\Omega/2\pi=2.75\,$GHz and a constant
$A$, corresponding to a horizontal slice in the presentations of \figs{figure2}{A}{D}.  The measured data (dots) in \fig{figure3}{A} feature a beating of broadened and overlapping current peaks.
The gray line is calculated for $\alpha_Z=1.5\cdot10^{-4}$ and $\lambda^\star=0$. Compared to our
measurement it shows a weaker broadening and a higher visibility. Much better
agreement is reached for $\lambda^\star=3.5\,\mu$eV (blue line).  This result
is robust under moderate variations of $\alpha_Z$ 
and does not depend on frequency or temperature.
\fig{figure3}{B} underlines the good agreement between measured (dots) versus calculated (lines) data by presenting $I(A)$ at $\bar\epsilon=n\hbar\Omega$ for various $n$ (vertical slices in \figs{figure2}{A}{D}). Owing to the electron-phonon interaction, the visibility of the interference pattern drops with increasing temperature (\fig{figure3}{C}).

To quantify $\alpha_Z$ with high accuracy, we use this temperature dependence
and thereby capture global information of the extended LZSM patterns (\figs{figure2}{A}{D}) by performing two-dimensional Fourier transformations $I(\bar\epsilon,A)\to\widehat I(\tau_\epsilon,\tau_A)$. The result, featured in \figs{figure2}{E}{H}, are simple, lemon-shaped structures of local maxima $\widehat I(\tau_\epsilon,\tau_A)\big|_\text{lemon}$.
Transforming Eq.~\eqref{LZSbasic} yields an analytic formula describing these lemon arcs:
\begin{equation}\label{eq:lemons}
\tau_A = \pm \frac{2k}{\Omega}
\sin\bigg(\frac{\Omega \tau_\epsilon + 2\pi k'}{2k}\bigg) ,
\end{equation}
with $k=1,2,3,\dots$, $k'=0,1,2,\dots$ and $k'<k$. Arcs for $k>1$ are a
consequence of $\Delta\gtrsim\gamma$, a prerequisite for observing a 
pronounced interference pattern (\suppl: \ref{dissipation_strength}). (Arcs for $k>1$ are weakly seen in
\figs{figure2}{E}{H}.  In superconducting qubits they have been also observed
but---considering $\Delta\ll\gamma$ \cite{Rudner2008a}---not explained.)
Concentrating on the principal lemon arc for $k=1$, we
find a non-monotonic behavior of $\widehat I(\tau_\epsilon,\tau_A)\big|_\text{lemon}$ with maxima at the arc's intersections (at
$\tau_A=0$ and $\tau_\epsilon$ a multiple of $2\pi/\Omega$).  Regions of decays
in-between have the form
\begin{equation}
\widehat I(\tau_\epsilon,\tau_A)\big|_\text{lemon}
\propto e^{-\lambda\abs{\tau_\epsilon}/\hbar}\,
        e^{-\frac{1}{2}(\lambda^\star\tau_\epsilon/\hbar)^2}\,,
\label{eq:Fourier}
\end{equation}
where the exponential term originates from the Lorentzian broadening due to
electron-phonon coupling and the Gaussian term describes the inhomogeneous
broadening caused by charge noise. Notice that $\tau_\epsilon$ is a Fourier
variable rather than a real time and, thus, $\lambda$ should not be interpreted
as physical decay rate.  (Only for $\Delta\ll\gamma$, all Lorentzians in
Eq.~\eqref{LZSbasic} possess the same width, so that $\widehat
I(\tau_\epsilon,\tau_A)\big|_\text{lemon}$ is described by Eq.~\eqref{eq:Fourier} with
simply $\lambda=\gamma$ as suggested in Ref.\ \cite{Rudner2008a}.)  In
\twofigs{figure4}{A}{C}
we plot measured and calculated decays (dots), respectively, for various
temperatures between 18\,mK and 500\,mK. The solid lines in panels A and C are
identical and express Eq.~\eqref{eq:Fourier} with $\lambda$ as fit parameter, while $\lambda^\star$ is kept fixed at $3.5\,\mu$eV.
\fig{figure4}{B} compares $\lambda(T)$ obtained by this procedure from our measurements (black dots) with the numerical results using three different values of $\alpha_Z$. An outstanding agreement between theory
and experiments is found at $\alpha_Z\simeq 1.5\cdot 10^{-4}$ (blue in
\fig{figure4}{B}).  This completes our set of model parameters needed to
calculate LZSM patterns as in \twofigs{figure2}{C}{D}.  $\lambda(T)$ increases
linearly for $T\gtrsim100\,$mK, while it is bounded by
$\lambda_\text{min}\simeq4\,\mu$eV at our lowest temperatures.
This bound marks the
intrinsic decay of $\widehat I(\tau_\epsilon,\tau_A)
\big|_\text{lemon}$ present even in
the low-temperature limit of our transport measurement but
is not related to the low temperature bound of the coherence time $T_2$.

To actually identify $T_2(T)$ we use its dependence on $\alpha_Z$ in the
spin-boson model. In the absence of rf-modulation, it provides the analytical
prediction \cite{weiss89,Makhlin2001b}:
\begin{equation}\label{eq:T2}
T_2(T,\alpha_Z)=
\frac{\hbar}{\pi\alpha_Z}\left( \frac{2\kT\bar\epsilon^2}{E^2}
+\frac{\Delta^2}{2E} \coth\left(\frac{E}{2\kT}\right)\right)^{-1}\,.
\end{equation}
In the low temperature limit, $\kT\ll E=\sqrt{\Delta^2+\bar\epsilon^2}$
our undriven qubit has, $T_2=2\hbar E/\pi\alpha_Z \Delta^2$.
Assuming $\bar\epsilon=0$ we find $T_2\simeq0.2\,\mu$s, which further increases at finite detuning. Alternatively, $T_2$ could be
increased by decreasing $\Delta$. This would, however,
reduce the clock-speed of the qubit.  In the same spirit, the rf-induced
renormalization of $\Delta\to\Delta_n$ stabilizes the qubit's coherence on the
expense of a larger gate operation time \cite{Fonseca2004a}.

Summarizing, we demonstrated that steady-state LZSM interferometry is a
viable tool to fully characterize a qubit including its coupling to a noisy
environment. The quantitative agreement between our experiments and our complete
system bath model analyzed with Floquet transport theory
allows us to trace the origins of inhomogeneous broadening and decoherence. Thereby we determined the individual values of $T_2^\star$ and $T_2$ of the qubit. Our steady-state method is remarkably simple compared to the alternative pulsed gate experiments. Our two-electron charge qubit is affected by slow charge noise limiting $T_2^\star=\hbar/\lambda^\star$ to $\simeq0.2\,$ns but a coherence time of $T_2\simeq0.2\,\mu$s being much longer than
previously reported values in quantum dot charge qubits
\cite{Hayashi2003a,Petersson2010a,Cao2013}. The clock-speed of our qubit, $\Delta/h\simeq3.1\,$GHz, which limits $T_2$ at $T\simeq20\,$mK and $\bar\epsilon=0$, would then provide enough time for $>600$ quantum operations. At higher temperatures or sizable $\bar\epsilon$ the decoherence is dominated by the electron-phonon coupling. Our method is simple, very general and can be applied to arbitrary qubit systems. An
extension including individually controlled Landau-Zener transitions and a
combination with non-adiabatic pulses will open up alternative means of quantum information processing.  Our two-electron qubit experiments illustrate an interesting approach for studying the interaction of qubits and complex many body quantum systems.

\subsection*{Acknowledgements}
We wish to thank R.\ Blattmann, E.\ Hoffmann, M.\ Kiselev, J.\ Kotthaus and P.\ Nalbach for
valuable discussions and are grateful for financial support from the DFG via
SFB-631 and the Cluster of Excellence \emph{Nanosystems Initiative Munich} and
by the Spanish Ministry of Economy and Competitiveness through Grant No.\
MAT2011-24331. S.\,L.\ acknowledges support via a Heisenberg fellowship of the
DFG.

\subsection*{Author contributions}
S.\,L.\ initiated and supervised the project. S.\,K.\
developed and implemented the theoretical model and performed the numerical
calculations. G.\,P.\ fabricated the DQD sample. G.\,P.\ and F.\,F.  performed
the experiments. D.\,S.\ and W.\,W.\ provided the wafer material. S.\,M.\ helped to develop the high frequency setup. S.\,K.,
G.\,P., F.\,F., P.\,H.\ and S.\,L.\ analyzed the data and wrote the manuscript.


\section*{METHODS}

\textbf{Experiment:}
The sample is based on a GaAs\,/\,AlGaAs heterostructure, grown by molecular
beam epitaxy, with a 2DES situated \SI{85}{\nano\meter} beneath the surface.
After the 2DES has been characterized at $T=\SI{4.2}\kelvin$, at which
temperature the carrier density is $n_\text{e}=\SI{1.19e11}{\per\centi\meter\squared}$ and the mobility is $\mu=\SI{0.36e6}{\centi\meter\squared\per\volt\per\second}$, we used
electron-beam lithography to fabricate the  nanostructure. The metal gates  in
\fig{figure1}{} contain 30\,nm of gold on top of 5\,nm of titanium. The DQD is
defined electrostatically by applying negative voltages to these gates and tuned
to the ground state configuration (1,1) (one electron per dot). The
inhomogeneous field of an on-chip nanomagnet partly lifts the Pauli-spin blockade
(detailed in reference \cite{Petersen2013a})
and allows LZSM measurements in the most interesting regime where charge and
spin qubits can coexist. Here we concentrate on a charge qubit and measure a
single-electron-tunneling dc current through the DQD while the voltage on one
gate is modulated at radio frequencies (\fig{figure1}{}). More details on the
DQD configuration, the measurements and the data analysis are provided in the
\suppl: \ref{suppl:experiment}.

\textbf{Theory:}
With our model we aim at realistic predictions for the dc current
$I(\bar\epsilon,A)$ through the DQD. It takes into account the nine most
relevant energy states for the charge configurations (1,0), (2,0), (1,1), and
(2,1) of the DQD Hilbert space, which are the states with chemical potentials
lying between or close to those of the source and drain leads. The Hamiltonian
includes inter-dot and dot-lead tunneling, inter- and intra-dot Coulomb
repulsion, Zeeman terms stemming from an inhomogeneous magnetic field and the
modulation of the on-site energies by applying an rf voltage to one gate. For
the coupling to the dissipative environment we employ a generalized spin-boson model.  To describe the periodically driven quantum system we use Floquet theory which is based on the ansatz $\psi_n(t) = e^{-i\omega_nt}\phi_n(t)$ with
$\phi_n(t)=\phi_n(t+2\pi/\Omega)$ resembling Bloch-functions but with the space
coordinate replaced by time, while the quasi-energy $\hbar\omega_n$ corresponds to
the quasi-momentum.  We then include the dot-lead tunneling and the action of
the dissipating environments within Bloch-Redfield theory. Using the Floquet
states as basis captures the influence of the rf-field and allows us to treat
the resulting master equation within a rotating-wave approximation. This offers
an important computational advantage because finding the steady-state solution
is now reduced to solving a linear, time-independent matrix equation, despite
the rf driving \cite{Kohler2005a}. Our experimental results do not indicate any
significant temperature dependence of the spin relaxation time (within the
temperature window explored).  Therefore, we treat the latter in a simplified
way using a Lindblad form with a phenomenological rate. Finally, we identify the
current operator: it corresponds to those terms of the master equation
describing the incoherent transition $(2,0)\to(1,0)$, i.\,e.\ the tunneling of
electrons from the left dot to the drain lead. The measured dc current
corresponds to the expectation value of the current operator. More details of
our theory are provided in the \suppl: \ref{suppl:model}.

\appendix
\section*{SUPPLEMENTARY INFORMATION}

In the main article we demonstrated that LZSM interferometry is a viable tool to
measure standard qubit properties and, beyond, to determine its coupling to a
noisy environment. In our specific case of a DQD charge qubit we found two main
noise sources: (i) slow environmental fluctuations resulting in an inhomogeneous
Gaussian broadening, and (ii) the heat bath, resulting in a homogeneous
Lorentzian broadening. In \app{suppl:experiment} we provide additional
experimental results together with numerical data, which underlie our
interpretations in the main article. Further, we detail our quantitative data
analysis based on a self-consistent fitting procedure and numerical calculations
resulting in the qubit-environment coupling constants, namely the standard
deviation $\lambda^\star=\hbar/T_2^\star$ of the inhomogeneous broadening and
the dimensionless dissipation strength $\alpha_Z$ of the coupling to the
phonons. 

In \app{suppl:model} we discuss the details of our model for the DQD and its
coupling to the leads as well as to the phonons.  Moreover, we sketch the
Bloch-Redfield master equation approach by which we compute the asymptotic state of the DQD and the time-averaged current.
This includes a discussion of how we extract the qubit's coherence time $T_2$
from $\alpha_Z$ and under which conditions this is an appropriate procedure.
Note that we neglect a second component of the electron-phonon coupling, namely
$\alpha_X$ which would mainly cause additional energy relaxation between the
qubit states. An estimate of $\alpha_X$ and a discussion, which justifies its
negligence, is provided in Appendix \ref{suppl:model}.

Table \ref{table} in \app{suppl:table} summarizes all system parameters extracted from various measurements and used for the numerical calculations.

\section{Additional Experiments and Data Analysis}\label{suppl:experiment}

\subsection{Initial tuning of the double quantum dot}\label{suppl:tuning}

\begin{figure*}
\begin{center}
\includegraphics[width=.85\textwidth]{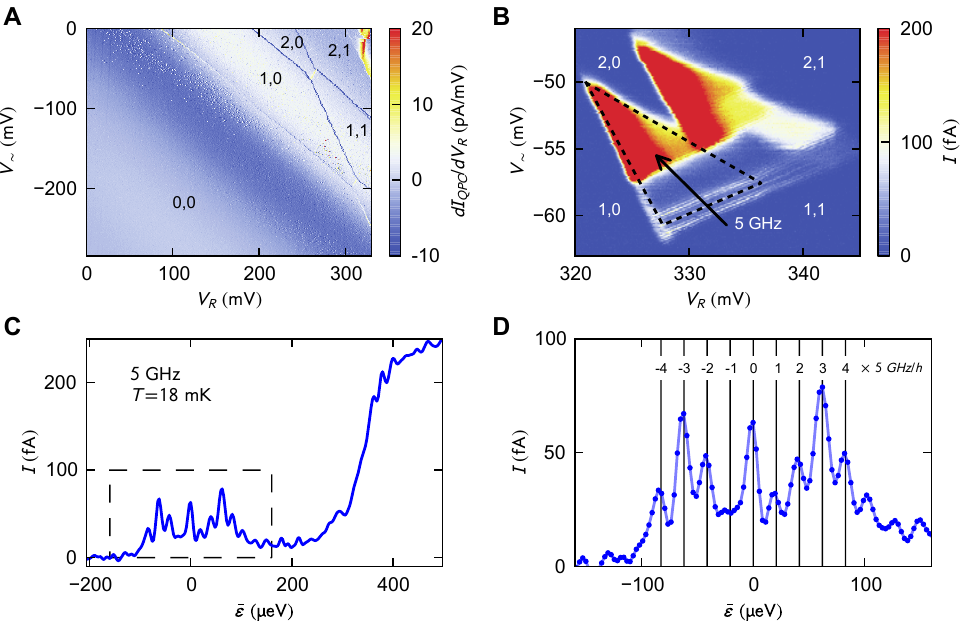}
\caption{\textbf{Initial tuning of the qubit.}
\textbf{(A)} Charge stability diagram of the unbiased DQD ($V=0$) as function of
dc gate voltages (the gates are marked in \fig{figure1}{A}),
measured by charge detection \cite{Field1993} at $T\simeq20\,$mK.  In detail, the
color scale displays the linear transconductance of a nearby almost pinched off
quantum point contact. It has been defined in the 2DES by $V_\sim$ and
the upper left gate (gray in \fig{figure1}{A}).  Sharp lines
of minimal transconductance are charging lines of the DQD which is empty
[configuration $(0,0)$] in the lower left half of the plot. 
\textbf{(B)} Current $I$ in the vicinity of the $(1,1)\leftrightarrow(2,0)$
transition of the charge stability diagram for $V=1\,$mV applied across the DQD
(see \fig{figure1}{A}) while $V_\sim$ was modulated with
frequency $\Omega/2\pi=5\,$GHz and amplitude $A\simeq80\,\mu$eV. The dashed
triangle marks the region of current via
$(1,0)\to(2,0)\leftrightarrow(1,1)\to(1,0)$. The black arrow indicates the
detuning axis $\bar\epsilon$ and $\bar\epsilon=0$ at its intersection with the
dashed line.
\textbf{(C)} $I(\bar\epsilon)$ measured along the black arrow in (B).
\textbf{(D)} Current around $\bar\epsilon=0$ corresponding to the region framed
by a dashed box in (C). The current maxima at $\bar\epsilon=n\hbar\Omega$ with
$n=0,\pm1,\pm2,\pm3,\pm4$ (vertical lines) are caused by PAT.
\label{fig:tuning}
}
\end{center}
\end{figure*}

Our experiments start by tuning the double quantum dot (DQD) by means of gate
voltages. As an orientation, \fig{fig:tuning}{A}
displays a charge stability diagram of the unbiased DQD ($V=0$) as function of
the gate voltages $V_\sim$ and $V_\text R$. It has been measured using a QPC as
charge detector. The sharp lines of local minima in transconductance $\text d
I_\text{QPC} / \text d V_R$ are the charging lines of the two dots which separate
regions of stable charge configurations ranging from $(0,0)$ to $(2,1)$.
\fig{fig:tuning}{B} details the region of the stability diagram near the
transition $(1,1)\leftrightarrow(2,0)$, but it plots the current $I$ measured
through the DQD as a response to $V=1\,$mV applied across the DQD (see
\twofigs{figure1}{A}{B}).
The finite current within the framed triangle is a consequence of the
single-electron tunneling cycle $(1,0)\to(1,1)\leftrightarrow(2,0)\to(1,0)$,
where the double arrow accounts for the fact that the interdot tunnel coupling
is coherent and large compared to the dot-lead tunnel couplings. The transition
$(1,1)\leftrightarrow(2,0)$ thereby divides into $\Slr\leftrightarrow
\Sll$ and $\Tlr\leftrightarrow \Tll$ while the coupling between singlet
and triplet subspaces is forbidden by the Pauli principle. This is the
configuration used for our LZSM interferometry measurements. The black arrow in
\fig{fig:tuning}{B} indicates the detuning axis. The current along this arrow, i.\,e.\ as a function of mean detuning $\bar\epsilon$, plotted in \fig{fig:tuning}{C}, shows two
interesting features: (i) $I$ is strongly suppressed for $\bar\epsilon<400\,\mu$eV,
because there the $\Tll$ state is beyond the transport window,
while the transition $\Tlr\to \Slr$ is hindered by Pauli-spin blockade
\cite{Ciorga2000a,Ono2002}, which makes $\Tlr$ a metastable state. In our case
the spin blockade is partly lifted especially near $\bar\epsilon=0$ where the
inhomogeneous field of our nanomagnet mixes $\Tlr$ and $\Slr$;
spin relaxation, provided by the hyperfine interaction with nuclear spins
\cite{Petersen2013a} also contributes, but is weaker. The strong current
increase at $\bar\epsilon\simeq400\,\mu$eV marks the onset of the $\Tll$
state contributing to the transport which then completely lifts the spin blockade via
the triplet channel $(1,0)\to \Tlr\leftrightarrow \Tll\to(1,0)$. (ii) In
\fig{fig:tuning}{B} we have, in addition,  applied an rf-modulation of $V_\sim$
at the frequency of 5\,GHz resulting in a modulation of the detuning with amplitude
$A\simeq80\,\mu$eV. This gives rise to a pattern of photon assisted tunneling
(PAT) current maxima appearing at $\bar\epsilon=n\hbar\Omega$ with
$n=0,\pm1,\pm2,\dots$. These PAT peaks in $I(\bar\epsilon)$, highlighted in
\fig{fig:tuning}{C}, transform into the LZSM patterns observed in our 2D plots
$I(\bar\epsilon,A)$.  Weaker PAT oscillations are also seen in panel C where the
$\Tll$-triplet starts to contribute to the current near
$\bar\epsilon\simeq400\,\mu$eV. (The actual singlet-triplet splitting in the
$(2,0)$-configuration is larger by the amplitude of
$A=80\,\mu$eV and accounts to $\simeq480\,\mu$eV.)

\begin{figure}
\begin{center}
\includegraphics[width=.95\columnwidth]{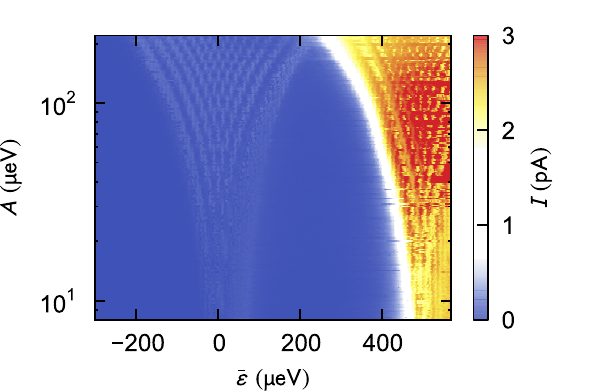}
\caption{\textbf{LZSM patterns: singlets versus triplets.}
$I(\bar\epsilon,A)$ at $\Omega/2\pi=2.5\,$GHz and $T\simeq20\,$mK (logarithmic amplitude axis).  Not only for the singlets transition $\Slr\leftrightarrow \Sll$ (left) but also for the triplets transition $\Tlr\leftrightarrow \Tll$ (right) is a LZSM interference pattern observed.
\label{fig:triplet}
}
\end{center}
\end{figure}

An example of measured LZSM interferometry is displayed in \fig{fig:triplet}{}
using a logarithmic $A$-axis. It clearly shows LZSM interference patterns involving both transitions $\Slr\leftrightarrow \Sll$ around $\bar\epsilon=0$ as well as  $\Tlr\leftrightarrow \Tll$ at larger $\bar\epsilon$ where the $\Tll$ state contributes to transport. In all other LZSM patterns presented in this article, the color scale is chosen to optimize the singlet contributions to the interference and the onset of the triplet channel is only seen as an asymmetry in $I(\bar\epsilon)$ at large $A$ (increased current in the upper right corner of e.\,g.\ \twofigs{figure2}{A}{B}).

\subsection{Energy calibration}\label{suppl:energy_calibration}

In this section we briefly explain how we determine the detuning $\bar\epsilon$ and the modulation amplitude $A$ from gate voltages, the source-drain voltage $V$ applied across the DQD and the modulation frequency
$\Omega$. The standard method is to use the current triangles in
\fig{fig:tuning}{B} which relate the known energy scale of the applied
source-drain voltage $eV$ to changes in gate voltages $V_\sim$ and $V_R$. The
relations are linear with the mutual gate-dot capacities as proportionality
factors \cite{Taubert2011a}. Here, we can refine such a standard calibration based
on the well known modulation frequency, which determines the LZSM interference
patterns, in the following way: (i) The current maxima appear at
$\bar\epsilon=n\hbar\Omega$ with $n=0,1,2,\dots$ which we use to calibrate
$\bar\epsilon(V_R,V_\sim)$. (ii) The positions of the minima of the current as
function of amplitude are also well known (see e.\,g.\ Eq.~\eqref{LZSbasic}) and we use
them to calibrate $A(V_R,V_\sim)$. At small frequencies, where the interference
patterns are less clear, the positions of the outermost current maxima (at
$A\simeq\bar\epsilon$) framing the region of finite current $I(\bar\epsilon,A)$
(e.\,g.\ in \twofigs{figure2}{A}{B}) can still be used for a
calibration. As the transmission of the rf modulation to the sample depends
on the frequency due to cable resonances in the experimental setup, the
calibration of $A$ has to be done separately for each frequency.

\subsection{Determination of the system parameters}
\label{suppl:couplings}

\begin{figure*}
\centering
\includegraphics[width=.84\textwidth]{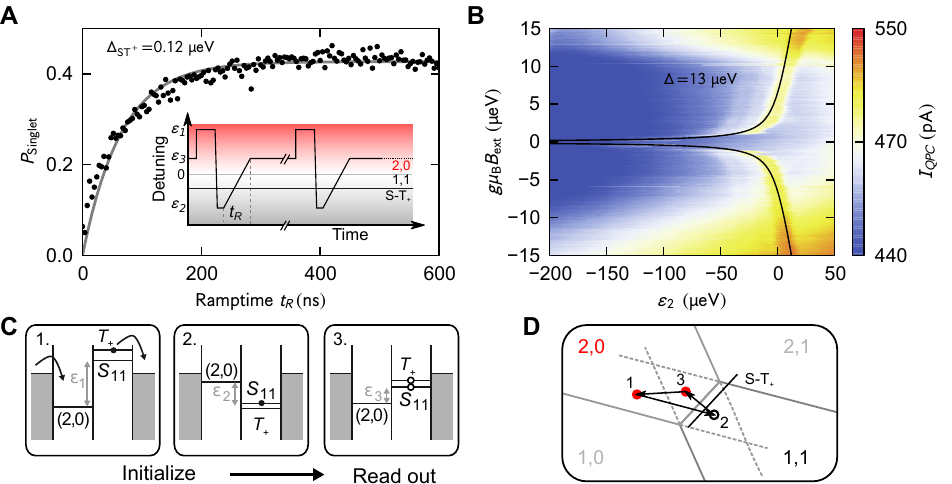}
\caption{{\bf Singlet-triplet and interdot tunnel couplings.} \textbf{(A)}~Probability to occupy the singlet state $\Slr$ after initialization in the same state and then sweeping across the $\Slr$-$T_+$ avoided crossing (at $T\simeq20\,$mK) as a function of the ramp time $t_\text R$ which is proportional to the inverse sweep speed $v$. From the theory (gray line) the singlet-triplet coupling is determined (details in bulk text). The inset sketches the pulse sequence.
\textbf{(B)}~Current $I_\text{QPC}$, measured at $T\simeq20\,$mK through the detector quantum point contact, as function of an external magnetic field and detuning $\epsilon_2$, where the entire pulse sequence shown in the inset of panel A is shifted vertically. Otherwise, the applied pulse sequence is the same as in  A, but with a quick and constant ramp time $t_\text R \simeq 1.1\,$ns. Enhanced current at the black lines (theory) mark singlet-triplet resonances. This so-called spin-funnel is used to determine the interdot-tunnel coupling $\Delta$ (details in bulk text).
\textbf{(C)}~Subfigures 1., 2., 3. sketch the chemical potentials (and dynamics) of the DQD at detunings $\epsilon_1$, $\epsilon_2$, $\epsilon_3$ in the inset of panel A.
\textbf{(D)}~Sketch of the DQD charge stability diagram indicating the positions at detunings $\epsilon_1$, $\epsilon_2$, $\epsilon_3$ during the pulse sequence sketched in the inset of panel A.
}
\label{fig:couplings}
\end{figure*}

Using our model we aim at a quantitative prediction of the measured current.
This requires knowledge of various system parameters such as the tunable tunnel
barriers and transition rates between triplet and singlet states. All parameters
used in our numerical calculations are summarized for convenience in Table
\ref{table} in Appendix \ref{suppl:table}. Following, we describe our
determination of those parameters, which are neither trivial nor described
elsewhere in this article. The largest energy scales are the intradot and
interdot Coulomb interactions $U\simeq3.5\,$meV and $U'\simeq0.8\,$meV. Knowing
the energy calibration (see last section) these values can be extracted from
charge stability diagram. In detail, $U$ corresponds to the distance between
charging lines in \fig{fig:tuning}{A} and $U'$ to the distance between the
triangle tips in \fig{fig:tuning}{B}. Next, we discuss the triplet-singlet
coupling, which in our case originates from the hyperfine interaction between
the electrons and many nuclei on the one hand and the inhomogeneous magnetic
field of our nanomagnet, shown in \fig{figure1}{A}, on the
other hand. The $\Tlr$ triplets split into
$T_+=\left|\uparrow\uparrow\right>$,
$T_0=\left(\left|\uparrow\downarrow\right>+\left|\downarrow\uparrow\right>\right)/\sqrt2$
and $T_-=\left|\downarrow\downarrow\right>$. The couplings between $T_+$ and
$\Slr$ and between $T_-$ and $\Slr$ are identical and caused by field
inhomogeneities parallel to the effective magnetic field (approximately parallel
to $B_\text{ext}=200\,$mT, see \twofigs{figure1}{A}{C}), while $T_0$ and
$\Slr$ are coupled by the perpendicular field inhomogeneities.  We actually
determine the $T_+$-$\Slr$ coupling by measuring the average charge
occupation in a continuously pulsed gate experiment. Here, we use a quantum
point contact as charge detector (while no voltage is applied across the DQD,
$V=0$).  As sketched in the inset of \fig{fig:couplings}{A} and in the energy
diagram in \fig{fig:couplings}{D},
we first initialize the DQD in $\Sll$ by applying a large positive detuning
$\epsilon_1$ (where the transition $\Tlr\to \Sll$ happens quickly via
charge exchange with the leads: $\Tlr\to (1,0) \to \Sll$; see left panel in \fig{fig:couplings}{C}). Next we prepare the DQD in the $\Slr$-state at $\epsilon=\epsilon_2$ by sweeping the detuning from  $\epsilon_1\to\epsilon_2$ at a constant speed obeying $\Delta^2\gg 2\hbar v/\pi\gg\Delta_{\text{ST},\pm}^2$. During this sweep, the DQD is adiabatically transferred from the $\Sll$ to the $\Slr$ state while passing the $\Sll$-$\Slr$ avoided crossing with coupling $\Delta$. The DQD also passes the $\Slr$-$T_+$ avoided crossing with coupling $\Delta_{\text{ST},\pm}$; this passage is, however, non-adiabatic because $\Delta_{\text{ST},\pm}\ll\Delta$ and, hence, the DQD remains in the $\Slr$-state. After waiting a short time
at $\epsilon_2$ (center panel in \fig{fig:couplings}{C}), we perform a
Landau-Zener passage within the ramp time $t_\text R$ through the $\Slr$-$T_+$ avoided crossing up to $\epsilon=\epsilon_3$ where we spend a relatively long
time in order to read out the charge state of the DQD (right panel in \fig{fig:couplings}{C}). We expect to find $(2,0)$
in case of a slow passage (with the DQD staying in the singlet subspace) and
$(1,1)$ in case of a fast passage bringing the DQD into the $T_+$ state, because
the decay $T_+\to \Slr\to \Sll$ is hindered by Pauli-spin blockade. \fig
{fig:couplings}{A} displays the probability to stay in the singlet subspace
$P_\text{singlet}=\alpha\left(1-P_\text{LZ}\right)=\alpha\left(1-\text{exp}\left[-\pi\Delta_{\text{ST},\pm}^2/2\hbar
v\right]\right)$ as a function of $t_\text R=(\epsilon_3-\epsilon_2)/v$. Fitting
this function (gray line in \fig {fig:couplings}{A}) to the measured singlet
probability indicates our $\Slr$-$T_+$ coupling of
$\Delta_{\text{ST},\pm}=(119\pm10)\,$neV, produced by our nanomagnet. The
pre-factor $\alpha\simeq 0.43$ accounts for the partial decay $T_+\to \Slr$
and depends on the duration of the readout period. Taking into account the interdot tunnel coupling which results in a reduced weight of $\Slr$ in the singlet eigenstate, this $\Delta_{\text{ST},\pm}$ corresponds to a magnetic field difference in the two dots giving rise to $g\mB\Delta B_x\simeq0.2\,\mu$eV, smaller than previously measured in the
same sample \cite{Petersen2013a}, which indicates a degradation by oxidation of
the single domain properties of our nanomagnet during six months of shelf
storage. Here, we used the g-factor $|g|=0.36$ as
determined for our DQD in Ref.\ \cite{Petersen2013a} and Bohr's magneton $\mB$. The hyperfine induced coupling contribution in our DQD is
$\Delta_\text{hyperfine}\simeq60\,$neV \cite{Petersen2013a} and results in a
corresponding inhomogeneous broadening of the singlet-triplet coupling. 

To determine the interdot tunnel coupling $\Delta$ we perform a so-called spin
funnel experiment \cite{Petta2005}. Thereby we repeat the same continuously
pulsed gate measurements as above but with a short and fixed $t_\text R=1.1\,$ns so
that all passages through the $\Slr$-$T_+$ crossing are now equally
non-adiabatic while the passage through the $\Slr$-$\Sll$ crossing is
still adiabatic (see inset of \fig{fig:couplings}{A}).  Under this conditions, a
pulse cycle $\epsilon_1\to\epsilon_3\to\epsilon_2\to\epsilon_3$ (\fig{fig:couplings}{D}) will usually
bring the system back to the $\Sll$ singlet after preparing the $\Sll$
singlet at the detuning $\epsilon_1$. A notable exemption occurs if $\epsilon_2$
coincides with the singlet-triplet resonance, namely for $\pm g\mB B_\text{ext}
= \frac 1 2\left(\pm\sqrt{\epsilon_2^2 + \Delta^2}-\epsilon_2\right)$, and if
the system spends sufficient time ($>h/\Delta_\text{ST}$) there to allow for a
singlet-triplet transition.
As a consequence, we measure a deviation from the $\Sll$ configuration at
this resonance during the readout at $\epsilon_3$, where the $\Tlr$ triplet
decays only slowly.  To map out this condition we plot in \fig{fig:couplings}{C}
the current $I_\text{QPC}$ through the detector QPC (corresponding to the
average charge state of the DQD) as a function of $g\mB B_\text{ext}$ and
$\epsilon_2$. The two distinct lines of enhanced $I_\text{QPC}$ correspond to a
finite occupation of one of the $T^\pm$ triplets.  By fitting the resonance
condition (above) we find our tunnel coupling $\Delta = (13 \pm 1)\,\mu$eV
(black lines in \fig{fig:couplings}{C}).

The initialization and decay rates $\Gamma_\text{in}$ and $\Gamma_\text{out}$,
respectively, are finally reconstructed from measuring the dc current through
the DQD as a function of detuning and as a function of source drain voltage in
forward and backward direction.  An example of such
a measurement is shown in \fig{fig:tuning}{C}, where in this case an
rf-modulation was applied in addition. Since the current in backward
direction (for $\mu_\text D>\mu_\text S$ in \fig{figure1}{B}) is practically independent of the spin relaxation, it allows us
to determine the dot-drain coupling $\Gamma_\text{out} = \Gamma_{L}$.  In
turn, the magnitude of the current in forward direction provides a faithful
estimate for $\gamma_\sigma$ and $\Gamma_\text{in}$.

\subsection{Origin of the inhomogeneous broadening}\label{suppl:lambdastar}

\begin{figure*}
\centering
\includegraphics[width=1.\textwidth]{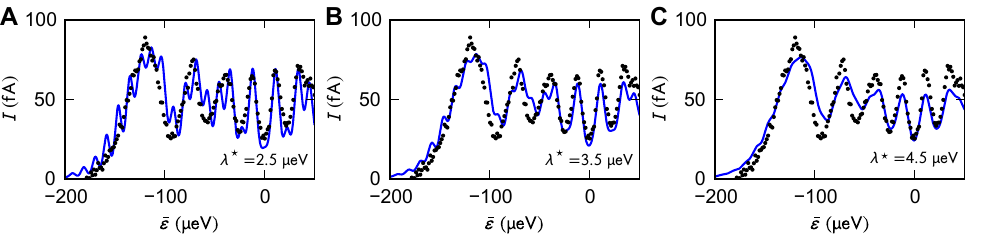}
\caption{\textbf{Accuracy in determination of inhomogeneous broadening.}
$I(\bar\epsilon)$ at constant amplitude $A=130\,\mu$eV and for $T=18\,$mK and $\Omega/2\pi=2.75\,$GHz as in \fig{figure3}{A}. The measured data (black dots) are identical in all three panels while the model curves (blue lines) are based on the same $\alpha_Z=1.5\times10^{-4}$ (our final value) but use various values of $\lambda^\star$. Best agreement with the measured data is reached for $\lambda^\star=3.5\,\mu$eV (panel B).}
\label{fig:differentalphas}
\end{figure*}

Compared to the measured LZSM patterns, our Floquet-Bloch-Redfield formalism,
which already takes into account the realistic electron-phonon coupling
$\alpha_Z$ (and the DQD parameters summarized in Table \ref{table} such as tunnel couplings), predicts a much higher
visibility of the interference pattern resulting in sharper current maxima.
This is evident in \fig{figure3}{A} which compares the
measured interference at a constant modulation amplitude with calculated data.
We resolved this caveat by introducing an additional Gaussian inhomogeneous
broadening $\lambda^\star$ (\fig{figure3}{A}), where the final values
$\alpha_Z=1.5\times10^{-4}$ and  $\lambda^\star=3.5\,\mu$eV have been calculated
self-consistently. \fig{fig:differentalphas}{}
demonstrates the convergence by plotting three calculated curves (lines) using various values of $\lambda^\star$ around its final value while the electron-phonon coupling is kept fixed at $\alpha_Z=1.5\times10^{-4}$ (as also in \fig{figure3}{A}). The model curve (blue line) in panel B using $\lambda^\star=3.5\,\mu$eV fits best to the measured data (dots).

The inhomogeneous broadening is a result of the combination
of slow charge noise and our time averaging dc measurement: The spectrum of
charge noise has been measured in heterostructures similar to ours. It can be
described as $1/f$-noise which typically occurs only at frequencies below
10\,kHz \cite{Fujisawa2000,Pioro-Ladriere2005,Buizert2008,Yacoby2013}. The
longest time scale of our experiment is the dwell time in the DQD of each
electron, contributing to the measured current. It is in the order of $1\,\mu$s,
much shorter than the highest frequency components of charge noise. A single
shot qubit measurement and hence $T_2$ is, consequently, unlikely to be affected
by charge noise. However, in our steady state experiments each measured data
point averages the dc current over $200\,$ms. Such an effective time ensemble
measurement can be inhomogeneously broadened by charge noise, slow compared to
$T_2$ but fast compared to the averaging time. Assuming a Markovian statistics,
this inhomogeneous broadening is well described using a Gaussian distribution
with standard deviation $\lambda^\star$.

\subsection{Dissipation strength}\label{dissipation_strength}

In comparison, determining the dissipation parameter $\alpha_Z$ requires
considerably more effort, experimentally and even more in theory, where it
enters in a rather complex manner.  We follow a route that is based on an idea by Rudner \textit{et al.} \cite{Rudner2008a}, who
showed analytically that the Fourier transformed $\widehat
I(\tau_\epsilon,\tau_A)$ of the dc current pattern exhibits a lemon-shaped
structure, composed of sinusoidal branches. However, the treatment of Ref.\
\cite{Rudner2008a} neglects the impact of the tunnel matrix element $\Delta$ on
the dynamical phase which finally yields an expression similar to our
Eq.~\eqref{LZSbasic}, but with $\Delta_n^2+\gamma^2$ in its denominator replaced by 
a phenomenological decay rate~$\widetilde\gamma^2$.
Thus, the result is a simple Lorentzian broadening of width $\widetilde\gamma$
giving rise to an exponential decay,
$\propto\exp(-\widetilde\gamma|\tau_\epsilon|)$, of the Fourier transformed
including the lemon structure. This simplification allows an analytical solution
of the problem for the price of limiting our horizon to an unrealistically weak inter-dot tunnel coupling and a convenient but just
phenomenologically introduced broadening.

In our DQD we have $\Delta_n\gtrsim\gamma$ for all relevant resonances
in the whole parameter range measured; more precisely the interdot tunnel coupling exceeds all broadening mechanisms including the initialization and decay rates,
$\Gamma_\text{in}$ and $\Gamma_\text{out}$, but also the broadening caused by
environmental influences. This guaranties a sufficiently long coherence time, $T_2>\hbar/\Delta$, which is a necessary condition for qubit operation as $\Delta/\hbar$ is the qubit clock-speed.

Interestingly, the finite $\Delta_n$ in the denominator of Eq.~\eqref{LZSbasic} has a
direct manifestation in the Fourier transformed of the measured LZSM patterns.
It gives rise to extra features described in Eq.~\eqref{eq:lemons} present in both our
measured and calculated data: cosine shaped arcs in the Fourier transformed
(marked by black arrows) in \figs{figure2}{E}{H} in addition to the main lemon
structure. These extra arcs are also evident in the measured data discussed in
reference \cite{Rudner2008a} but they have not been reproduced in the
calculations there (for the reasons discussed above).

\begin{figure*}
\centering
\includegraphics[width=.9\textwidth]{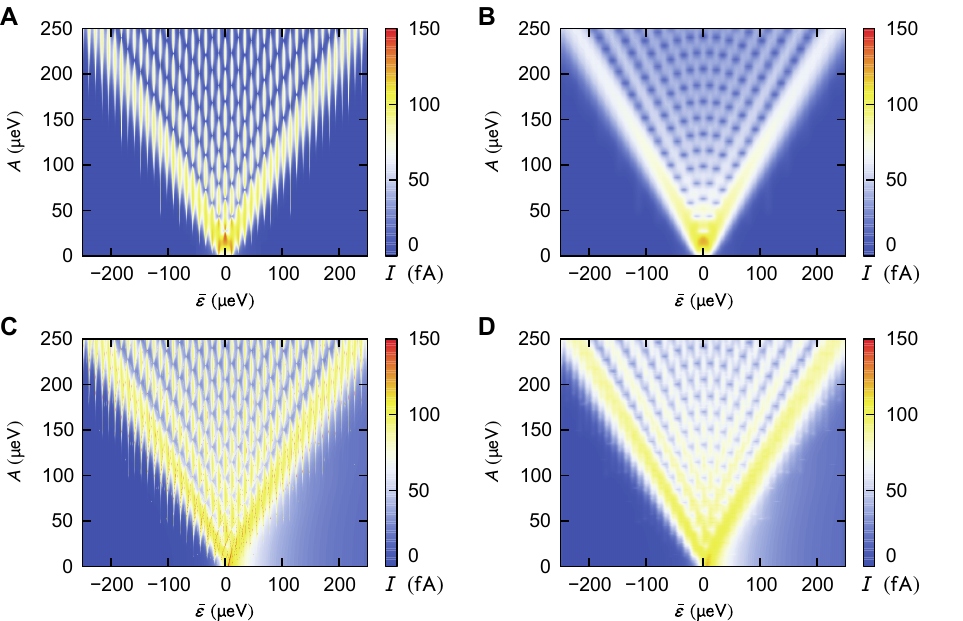}
\caption{\textbf{Eq.~\eqref{LZSbasic} versus full model.}
All data are calculated using $\Omega/2\pi=2.75\,$GHz and system parameters as listed in Table \ref{table}.
(\textbf A) Analytical solution of Eq.~\eqref{LZSbasic}. The width of the
current peaks as function of $\bar\epsilon$ is
$\delta\bar\epsilon=\sqrt{\Delta_n^2+\gamma^2}$ (containing only tunneling
rates), see Eq.~\eqref{LZSbasic}.  
(\textbf B) Analytical solution of Eq.~\eqref{LZSbasic}, as in panel A, but additionally convoluted with a Gaussian profile of width $\lambda^\star=3.5\,\mu$eV to simulate the effect of an inhomogeneous broadening in a time ensemble measurement caused by slow charge noise.
(\textbf C) Numerical solution of our full model, including electron-phonon coupling, at $T=18\,$mK (with parameters from Table \ref{table}) but using $\lambda^\star=0$.
(\textbf D) Numerical solution of our full model as in panel C but using $\lambda^\star=3.5\,\mu$eV. The similarity between the analytical solutions and the low temperature result of our full model, if comparing panel A with C and panel B with D, justifies our perturbative approach to treat the electron-phonon coupling.
}
\label{fig:pat}
\end{figure*}

\begin{figure*}
\centering
\includegraphics[width=.8\textwidth]{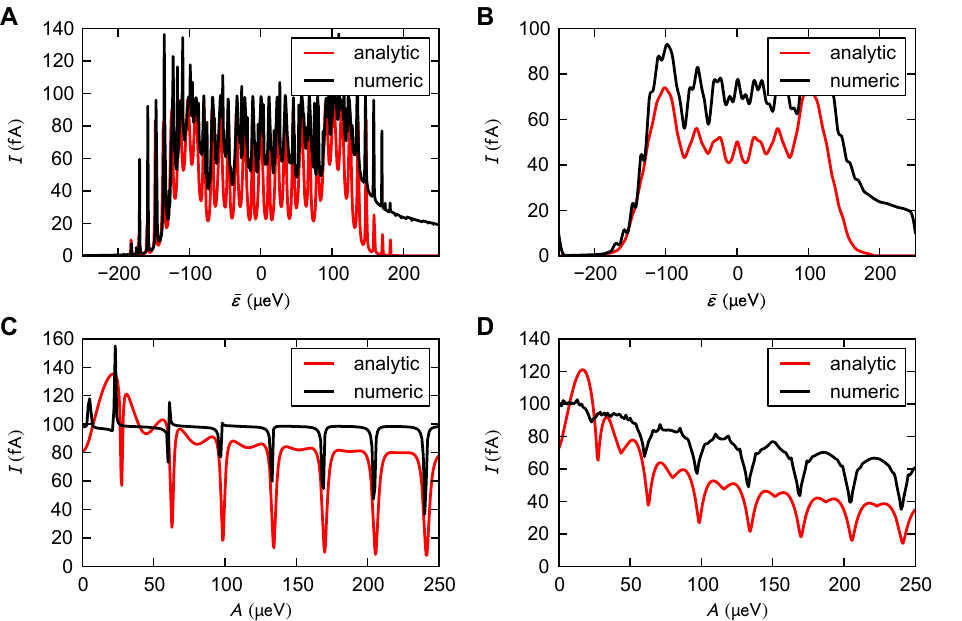}
\caption{\textbf{Eq.~\eqref{LZSbasic} versus full model --- slices.}
Slices of the data in \fig{fig:pat}. Each panel compares the analytical
predictions of Eq.~\eqref{LZSbasic} (marked as ``analytic'') with those of our full model for $\alpha_Z=1.5\times10^{-4}$ and at $T=18\,$mK (marked ``numeric''). The modulation frequency is $\Omega/2\pi=2.75\,$GHz.
(\textbf A) $I(\bar\epsilon)$ at $A\simeq130\,\mu$eV using $\lambda^\star=0$.  
(\textbf B) $I(\bar\epsilon)$ at $A\simeq130\,\mu$eV using $\lambda^\star=3.5\,\mu$eV.
(\textbf C) $I(A)$ for $\bar\epsilon=0$ using $\lambda^\star=0$.  
(\textbf D) $I(A)$ for $\bar\epsilon=0$ using $\lambda^\star=3.5\,\mu$eV.  
For the agreement of the full model results with measured data, see
\twofigs{figure3}{A}{B} of the main text.
\label{fig:pat2}
}
\end{figure*}

The analytical expression in Eq.~\eqref{LZSbasic} serves as a sign post for our
analysis as it describes the main features of our measurements correctly. This is evident in \fig{fig:pat}{} and \fig{fig:pat2}{}.
which provide a direct comparison between the predictions of
Eq.~\eqref{LZSbasic} and our full model. The detailed comparison between our
numerical calculations and measurements, provided in \twofigs{figure3}{A}{B},
further demonstrates that our full model fits considerably better to our data
than Eq.~\eqref{LZSbasic}.
The analytical expression in Eq.~\eqref{LZSbasic} only considers non-interacting electrons and, hence, fails to predict decoherence
effects. A reliable physical interpretation
including the observed temperature dependence (see Figs.\ \ref{figure3}C and \ref{figure4} requires a detailed analysis: First, it is necessary to explicitly
consider all (dot-lead and interdot) tunnel couplings and the relevant energy
spectrum of the DQD.  Second, interaction effects have to be included which in
our case comprise: (i) Coulomb-interaction giving rise to Coulomb blockade
and the coupling to charge noise; (ii) exchange interaction causing Pauli-spin
blockade, hyperfine interaction causing spin-flips and the mixing between
singlet and triplet states by the inhomogeneous field of the nanomagnet; (iii)
electron-phonon interaction resulting in decoherence. We focus on the latter.
Our master equation formalism takes into account all these effects and allows us
to numerically calculate $I(\bar\epsilon,A)$ in the range in which we take our
experimental data and compute its two-dimensional discrete Fourier
transformation.  Finally, a Fourier transformation of our data causes cut-off
effects, because both measured and calculated data span only finite ranges in
$\bar\epsilon$ and $A$. Typical artefacts of the discrete Fourier transformation are avoided throughout
our analysis as good as possible by using only data with sufficiently high
resolution.

\begin{figure*}
\includegraphics[width=.9\textwidth]{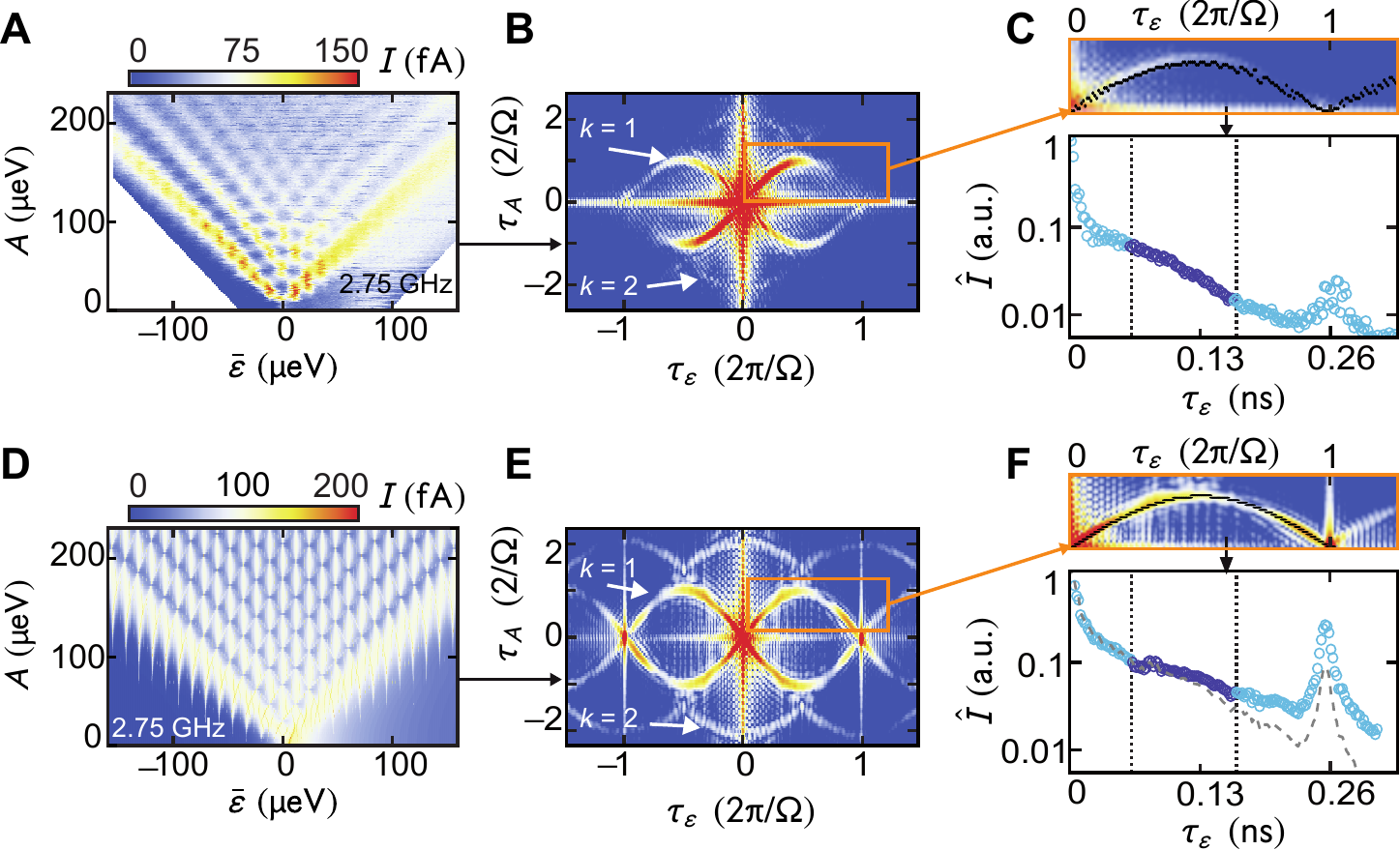}
\caption{\textbf{Data analysis based on Fourier transformation.}
Analysis steps on measured data (\textbf A--\textbf C) versus theory (\textbf D--\textbf F) for $T\simeq60\,$mK and $\Omega/2\pi=2.75\,$GHz using the parameters listed in Table \ref{table} for the numerical calculations. The numerical data in panels D--F are without the inhomogeneous broadening, i.\,e.\ $\lambda^\star=0$.
(\textbf A and \textbf D) LZSM interference patterns $I(\bar\epsilon,A)$. The measured data in panel A display a stronger asymmetry in $\bar\epsilon$ compared to the numerical data in panel D, which is discussed in \sect{suppl:DNP}.
(\textbf B and \textbf E) Two-dimensional Fourier transformed $\widehat
I(\tau_\epsilon,\tau_A)$ of the raw data in panels A and D. Clearly visible are
the principal lemon arcs for $k=1$ and those for $k=2$ in Eq.~\eqref{eq:lemons}. Horizontal and vertical lines at $\tau_\epsilon=0$ and $\tau_A=0$, respectively, are artefacts caused by the discrete Fourier transformation. The color scales are in arbitrary units as the absolute amplitude of $\widehat I$ scales with the number of data points in the raw data and has no physical meaning. The higher visibility of the numerical data in panel E compared to the measured data in B is due to the negligence of the inhomogeneous broadening. 
(\textbf C and \textbf F) Decay of a quarter of the principal lemon arc in the
range $0\lesssim \tau_\epsilon \lesssim 2\pi/\Omega$. The dashed line in the
lowest panel is the data (open circles) multiplied by the Gaussian
$\exp[-\frac12(\lambda^\star\tau_\epsilon/\hbar)^2]$ using
$\lambda^\star=3.5\,\mu$eV for direct comparison with the data in panel C. The
region between the two vertical (dotted) lines is then fitted with Eq.~\eqref{eq:Fourier} to determine $\lambda$.  
}
\label{fig:analysis}
\end{figure*}

Next, we discuss the details of our data analysis: after calculating raw data resembling the measured LZSM patterns, using estimated values for $\alpha_Z$ and $\lambda^\star$ we apply the identical analysis to experimental and numerical data. Then, we compare the results and repeat calculation and analysis of the numerical data with modified $\alpha_Z$ and $\lambda^\star$ in a self-consistent way until we find best agreement with the measured data. \fig{fig:analysis}
demonstrates the last step of this procedure, exemplarily, for a typical set of measured data
in panels A--C and the corresponding calculated data, based on our model parameters listed in Table \ref{table}, in panels D--F. The numerical data shown here neglect the inhomogeneous broadening, equivalent to using $\lambda^\star=0$, as this is sufficient for evaluating $\lambda$ from the numerical data. Measured and calculated data in \fig{fig:analysis}, therefore, imply differences (details in figure caption); for a direct comparison including the inhomogeneous broadening in the numerical data we refer to \fig{figure2}{} and \fig{figure3}{}.
Panels A and D  of \fig{fig:analysis}{} show the measured and calculated
$I(\bar\epsilon,A)$, respectively, panels B and E the corresponding Fourier
transformed $\widehat I(\tau_\epsilon,\tau_A)$.
Since the current is real-valued, the Fourier transformed pattern is point
symmetric, while the approximate mirror symmetry at the $A$-axis relates the
two independent branches.  Therefore it is sufficient to restrict the analysis
to the upper-right quarter as indicated in \twofigs{fig:analysis}{C}{F}.
The Fourier transformed of the current along the
lemon arcs $\widehat I(\tau_\epsilon,\tau_A)\big|_\text{lemon}$ incorporate a
decay between two maxima at the arcs intersections at $\tau_\epsilon=0$ and
$\tau_\epsilon=2\pi/\Omega$. These maxima indicate a fast intrinsic decay
and are related to $\Delta_n^2$ dominating the denominator of Eq.~\eqref{LZSbasic} near the $n$-photon resonances. They are not a measure of the qubit decoherence. Note that finite range cut-off effects of the Fourier
transformations cause the finite $\widehat I(\tau_\epsilon,\tau_A)$ along
$\tau_\epsilon=0$ and $\tau_A=0$ in \twofigs{fig:analysis}{B}{E}, which
additionally obscure the maxima in $\widehat
I(\tau_\epsilon,\tau_A)\big|_\text{lemon}$. For our further analysis we
therefore only consider the decay of $\widehat I(\tau_\epsilon,\tau_A)\big|_\text{lemon}$in the regions marked in \twofigs{fig:analysis}{C}{F}.
To determine $\lambda$ the measured $\widehat
I(\tau_\epsilon,\tau_A)\big|_\text{lemon}$ in \fig{fig:analysis}{C} is fitted
with Eq.~\eqref{eq:lemons} using $\lambda^\star=3.5\,\mu$eV while the calculated
data in \fig{fig:analysis}{F} are just fitted with the exponentially decaying
term  in Eq.~\eqref{eq:lemons} using $\lambda^\star=0$.

\begin{figure*}
\includegraphics[width=\textwidth]{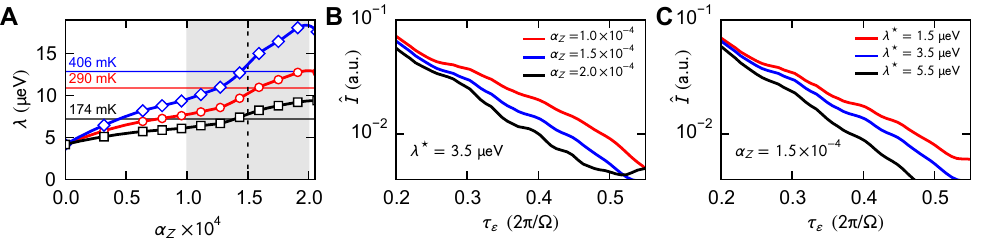}
\caption{\textbf{Analysis of the Fourier transformed --- electron-phonon coupling.}
(\textbf{A}) Decay rate $\lambda(\alpha_Z)$ for three different temperatures. The curves are numerically calculated using the fixed $\lambda^\star=3.5\,\mu$eV but various values of $\alpha_Z$. Horizontal lines indicate $\lambda$ determined from our measured data for the same three temperatures. The vertical dashed line and the gray region indicate the best fitting $\alpha_Z$ and its accuracy. Also compare to \fig{figure4}{B}.
(\textbf{B}) Numerically calculated decay of the principal lemon arc for $T=295\,$mK using the fixed $\lambda^\star=3.5\,\mu$eV but various values of $\alpha_Z$ (same as those in \fig{figure4}{B}). 
(\textbf{C}) Numerically calculated decay of the principal lemon arc for $T=295\,$mK using the fixed $\alpha_Z= 1.5\times10^{-4}$ but various values of $\lambda^\star$. 
}
\label{fig:decay_model}
\end{figure*}

To accurately determine the electron-phonon coupling we consider the temperature dependence of $\lambda(\alpha_Z,T)$ rather than relying on a single LZSM pattern at low temperature. This procedure allows us to properly separate the two main noise sources, the temperature independent charge noise giving rise to the inhomogeneous dephasing time $T_2^\star$ and the temperature dependent homogeneous broadening $\lambda$, which is directly related to $\alpha_Z$ and determines the qubit decoherence time $T_2$. (There are no indications for a temperature dependence of the charge noise for $T<1\,$K). The details of this procedure are discussed in the main article around \fig{figure4}.
In \fig{fig:decay_model}{A}
we provide additional data on the determination of the electron-phonon coupling
presenting $\lambda(\alpha_Z)$ for various temperatures. Each data point has
been determined from fitting Eq.~\eqref{eq:Fourier} to a principal lemon arc as the one in \fig{fig:analysis}{F} calculated using the fixed $\lambda^\star=3.5\,\mu$eV but various values of $\alpha_Z$. Horizontal lines indicate $\lambda$ determined from our measured data for the same three temperatures. Assuming a temperature independent $\alpha_Z$ we find best agreement to our data for $\alpha_Z = (1.5 \pm 0.2) \times 10^{-4}$ (vertical dashed line and gray region). Note that a similar information is contained in \fig{figure4}{B}.
The saturation of $\lambda(T)$ for low temperatures at a value
$\lambda_\text{min}\simeq4\,\mu$eV is a consequence of measuring PAT current peaks which possess the intrinsic width $\sqrt{\Delta_n^2+\gamma^2}$ of $I(\bar\epsilon,A)$ as a function of
$\bar\epsilon$ expressed in Eq.~\eqref{LZSbasic}. As is evident from \fig{figure4}{B}
of the main article the lower bound $\lambda_\text{min}$ is also observed in our experiments.
\twofigs{fig:decay_model}{B}{C} demonstrate the robustness of our main model parameters $\alpha_Z$ and $\lambda^\star$, respectively, by varying each of the two parameters separately and comparing the decay of the principal lemon arcs.

\subsection{Summary of data analysis}\label{suppl:analysis_summary}

Summarizing our data analysis, we started by determining all important physical constants such as tunnel couplings and spin-flip rates based on a number of independent measurements on our double quantum dot device already tuned to the configuration used for the LZSM interferometry experiments. To determine the remaining key-parameters $\lambda^\star$ and $\alpha_Z$ we used a self-consistent approach within our model. It turned out that $\alpha_Z$ could be best determined from the two-dimensional Fourier transformed of LZSM interference patterns at various temperatures. In contrast, $\lambda^\star$, which causes a strong but temperature independent inhomogeneous broadening, could be equally well determined from the raw data. This allowed us to avoid a third fit parameter (besides $\lambda$ and a prefactor) by which we would loose precision in finding $\alpha_Z$.
Specifically, by comparing the resulting decay rates $\lambda(T,\alpha_Z)$ with
those extracted from our measurements (see \fig{figure4}{B}), we find that in
our setup decoherence can be described by a Caldeira-Leggett model with Ohmic
spectral density and the dimensionless dissipation strength $\alpha_Z
\simeq1.5\times10^{-4}$.

\subsection{LZSM interference at various frequencies}\label{suppl:frequencies}

\begin{figure*}
\begin{center}
\includegraphics[width=1\textwidth]{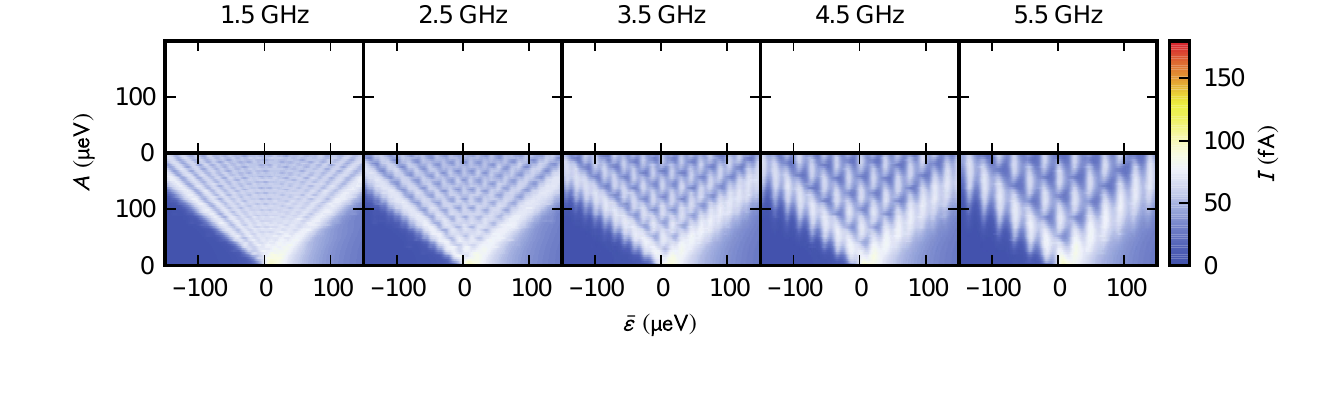}
\caption{\textbf{LZSM interference patterns for various modulation frequencies.} The upper row
contains measured and the lower row calculated data, both for $T\simeq20\,$mK. Model parameters according to Table \ref{table}. 
\label{fig:frequencies}
}
\end{center}
\end{figure*}

In the main article we already demonstrate that the LZSM interference pattern
depends on frequency. In \fig{fig:frequencies}{} we extend the frequency range presenting data between
$1.5\,\text{GHz}\le\Omega/2\pi\le5.5\,$GHz all measured (upper line) or numerically calculated (lower line) at $T\simeq20\,$mK.
At the highest frequencies we observe clear PAT patterns as expected from
Eq.~\eqref{LZSbasic} which distort increasingly as the frequency is lowered and
neighbored PAT current peaks overlap. At $\Omega/2\pi=1.5\,$GHz all interference
signatures are (almost) lost as $\hbar\Omega\simeq6\,\mu\text{eV}$ is close to the broadening caused by the combination of our $\lambda^\star\simeq3.5\,\mu$eV and $\lambda(T=20\,\text{mK})\simeq4\,\mu$eV.

It is instructive to estimate the Landau-Zener probability
$P_\text{LZ} = \exp(-\pi\Delta^2/2\hbar|v|)$ \cite{Landau1932a,Zener1932a,
Stueckelberg1932a, Majorana1932a} for these frequencies for an intermediate amplitude, say
$A=100\,\mu$eV.  For $\epsilon(t) = \bar\epsilon+A\cos(\Omega t)$, the sweep
velocity at the avoided crossing is $|v|= \Omega\sqrt{A^2-\bar\epsilon^2}$.
Thus we find for the frequencies used in \fig{fig:frequencies}{} and
$\bar\epsilon=0$ Landau-Zener transition probabilities in the range
$0.65\lesssim P_ \text{LZ}\lesssim0.89$ and $P_ \text{LZ}\simeq0.79$ for
$\Omega/2\pi=2.75\,$GHz where we performed our temperature dependent
measurements.  For non-vanishing $\bar\epsilon$, $P_\text{LZ}$ is smaller, so
that the average over all relevant crossings becomes of order
$P_\text{LZ}\simeq 1/2$ which ensures good visibility.  For frequencies
$\Omega/2\pi \lesssim 2\,$GHz,
the analytic estimate of Eq.~\eqref{LZSbasic} based on PAT becomes increasingly inaccurate and, consequently, our interpretation of the lemon
arc decay given by Eq.~\eqref{eq:Fourier} is not guaranteed.

\subsection{Influence of dynamic nuclear polarization}\label{suppl:DNP}

The measured data, e.\,g.\ in \twofigs{figure2}{A}{B} and in
\figs{fig:frequencies}{A}{E} contain two distinct features not included in our
model. The first one is a pronounced asymmetry in $I(\bar\epsilon)$ in the limit
of large amplitudes, which is considerably smaller in the theoretical data, which neglect the influence of the $\Tll$ triplet. The stronger asymmetry observed in measured data is indeed caused by the influence of the $\Tll$ state which grows with increasing positive detuning.
The effect is clearly seen in \fig{fig:triplet}{} and has been discussed at the end
of Sec.\ \ref{suppl:tuning}. The second feature occurs at very small amplitudes
and appears as if the tip of the current triangle was shifted to slightly
positive values of $\bar\epsilon$. It is a signature of dynamic polarization of
the nuclear spins caused by the hyperfine interaction between the current
carrying electrons and the nuclear spins in the DQD. At very small $A$, the
rf-modulation is practically off and we simply measure the current through the
DQD while sweeping $\bar\epsilon$ from positive towards negative values. As
explained in detail in Ref.\ \cite{Petersen2013a}, the current maximum occurs at
the value of $\bar\epsilon$ that marks the resonance between the $T_-$ and the
singlet state. This resonance is shifted towards positive $\epsilon$ by dynamic nuclear
polarization \cite{Petersen2013a}. The fact that the shift only occurs at very
small $A$ indicates that the continues rf-modulation effectively prevents the
polarization of nuclear spins. We therefore, do not have to include this in
our model here as long as we concentrate on data with $A>20\,\mu$eV.

\subsection{Temperature dependence and limitations of our model}\label{suppl:temperature}

\begin{figure*}
\begin{center}
\includegraphics[width=1\textwidth]{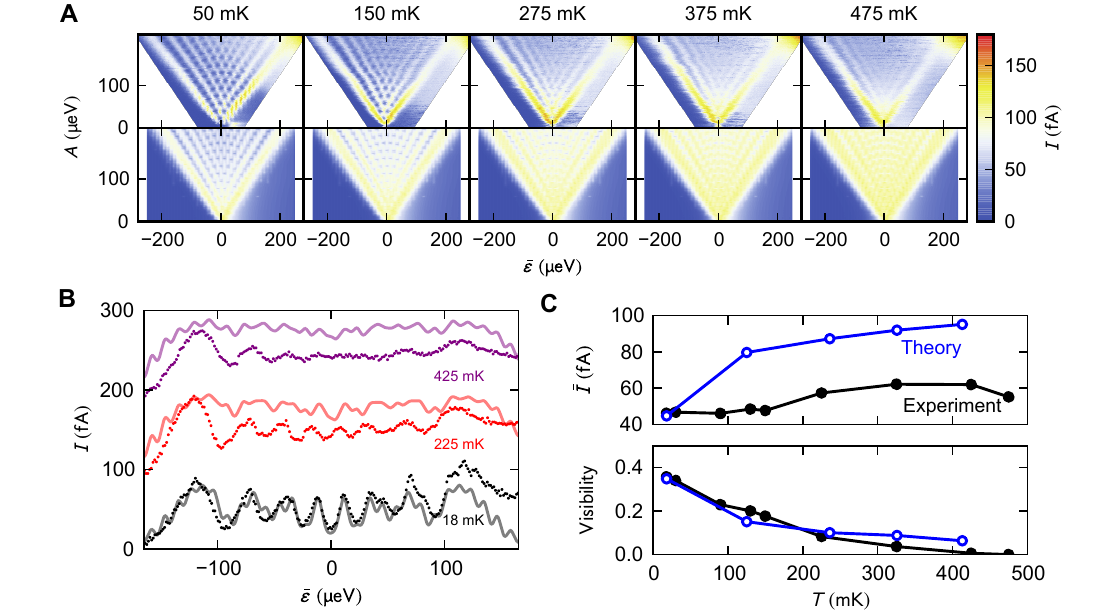}
\caption{\textbf{Temperature dependence of raw data versus model.}
(\textbf A) LZSM interference patterns at $\Omega/2\pi=2.75\,$GHz and various temperatures. The upper plots have been measured, the lower plots calculated using our full model with the parameters listed in Table \ref{table}. 
(\textbf B) Horizontal slices from plots like those in panel A at $A=130\,\mu$eV. Dots correspond to measured and lines to numerical data. 
(\textbf C) Average current $\bar I$ (upper panel) and visibility $\nu=\left(I_\text{max}-I_\text{min}\right)/\left(I_\text{max}+I_\text{min}\right)$ (lower panel), extracted from the slices presented in panel B and similar data, as function of temperature. The data points are taken in the region $-100\,\mu\text{eV}\le\bar\epsilon\le30\,\mu\text{eV}$; this avoids the region of $\bar\epsilon>100\,\mu$eV where experimental data are influenced by the $\Tll$ triplet which is neglected in our model. Black dots are measured while blue open circles are calculated.
}
\label{fig:temperature}
\end{center}
\end{figure*}

The temperature dependence of the LZSM patterns discussed in the main article in the Figures \ref{figure3}C and \ref{figure4} is at the heart of our model as it is used to accurately determine the electron-phonon coupling and finally the qubit coherence time $T_2$. So-far we have mostly concentrated on the temperature dependence of the principal lemon arc in the Fourier transformed of the LZSM patterns. \fig {fig:temperature}{A}
shows LZSM patterns measured (upper line) versus calculated (lower line) at
various temperatures. \fig {fig:temperature}{B} shows horizontal slices at
$A=130\,\mu$eV (dots are measured and lines present numerical data). To
facilitate a quantitative comparison, we extract the visibility as well as the
average current $\bar I$ in the region defined by
$-70\,\mu\text{eV}\le\bar\epsilon\le70\,\mu$eV and plot their temperature
dependences of experimental versus numerical data in \fig {fig:temperature}{C}.
The temperature dependence of the calculated visibility resembles the measured
ones. However, the measured mean current is roughly temperature independent
while the predicted mean current increases with temperature. Interestingly, the
decay of the principal lemon arc of the Fourier transformed is strongly related
to the visibility of a LZSM pattern but not at all to the mean current. Our
model, therefore, describes the decoherence of the qubit correctly while it
predicts an increase of the mean current with temperature that is stronger
than in the measurements.
A possible explanation for this discrepancy is that in the experiment, the
spectral density of the phonons is not strictly ohmic as is assumed in our
model, see Sec.~\ref{suppl:system-bath_hamiltonian}. Consequently, our
theoretical description may overestimate the thermal activation of some
singlet-triplet transitions.

\section{Theoretical modelling}\label{suppl:model}

Our aim is to compute the LZSM interference patterns using a realistic
model and to compare the results with the measured ones. Therefore we need to consider besides our DQD also its coupling
to electron source and drain contacts, as well as to environmental fluctuations.
In order to realistically describe our measurements, performed in the regime of Pauli-spin
blockade which is partly lifted by an inhomogeneous Zeeman field of an on-chip
nanomagnet, we further include spin relaxation. Our model considers all
energetically accessible DQD states and all processes which play a noticeable
role in our experiment. Comparing
our theoretical and experimental data we find two main contributions to a noisy
environment: the first is slow charge noise
\cite{Fujisawa2000,Pioro-Ladriere2005,Buizert2008,Yacoby2013} and can be described as an inhomogeneous broadening $\lambda^\star$, the second is the heat bath and contributes via the electron-phonon coupling $\alpha_Z$.
It is straightforward to independently extract other relevant system parameters from transport measurements, so that $\lambda^\star$ and $\alpha_Z$ are the only fit parameters left.

\subsection{System-lead-bath model}\label{suppl:system-lead-bath}

\subsubsection{Double quantum dot Hamiltonian}\label{suppl:hamiltonian}

We include the single-particle energies $\epsilon_L$ and $\epsilon_R$ in the
left\,/\,right dot, the electron-electron interactions neglecting the small
exchange terms, inter-dot tunneling, and the inhomogeneous Zeeman field to
obtain in second quantization the DQD Hamiltonian
\begin{equation}
\label{H_DQD}
\begin{split}
H_\text{DQD} = {} &
       \sum_{\ell=L,R} \epsilon_\ell n_\ell
      + U\sum_{\ell=L,R} n_\ell(n_\ell-1) + U'n_L n_R
\\  &      + \frac{\Delta}{2\sqrt{2}}\sum_{m=\uparrow,\downarrow}
           (c_{Lm}^\dagger c_{Rm} +c_{Rm}^\dagger c_{Lm})
\\ &
 +\frac{g\mB}{2} \sum_{\ell=L,R} (c_{\ell,\up}^\dagger,c_{\ell,\down}^\dagger)
       (\vec B_\ell\cdot\vec\sigma) (c_{\ell,\up},c_{\ell,\down})^T
,
\end{split}
\end{equation}
where $n_\ell = \sum_{m=\ud} c_{\ell m}^\dagger c_{\ell m}$ is the
occupation of dot $\ell=L,R$ expressed with the usual fermionic
creation and annihilation operators.
The largest energy scales are the intra- and inter-dot Coulomb interactions $U$
and $U'$, which define the diabatic basis states of our charge qubit with
energies $\epsilon(\Sll)=2\epsilon_L+U$ and
$\epsilon(\Slr)=\epsilon_L+\epsilon_R+U'$ and their mutual detuning
$\bar\epsilon\equiv\epsilon(\Sll) -\epsilon(\Slr)
=\epsilon_L-\epsilon_R+U-U'$.  The fourth term describes tunneling between the
dots with the matrix element $\Delta$ defined such that it equals the energy
splitting of the charge qubit formed by the singlets.  The final Zeeman term
affects the triplet states and, because of the inhomogeneous field contribution
of our on-chip magnet, also mixes singlets with triplets.  This mixing enables
transitions between singlet and triplet states and may be rather sensitive
to thermal exitations \cite{Ribeiro2013b}. If a source-drain
voltage is applied across the DQD, it causes a finite average current.  Notice
that also the hyperfine interaction causes electron
spin-flips, which we capture by a phenomenological spin-flip rate~$\gamma_\sigma$.

\subsubsection{Dot-lead Hamiltonians}

To model the single electron tunneling current through the DQD we have
to consider its interaction with the two-dimensional leads.  Starting
from the configuration (1,0), the right quantum dot is loaded via the
tunneling process $(0,1)\to(1,1)$ from the source contact, i.\,e.\ the
right lead. The latter is modeled as non-interacting electrons with
the Hamiltonian
$H_\text{lead} =
\sum_{q,m=\up\down} E_q c_{q,m}^\dagger c_{q,m}$ while the dot-lead coupling terms reads
\begin{equation}
\label{leadcoupling}
H_\text{dot--lead} = \sum_{q,m=\ud} V_{q}(c_{R,m}^\dagger c_{q,m}
+c_{q,m}^\dagger c_{R,m}) .
\end{equation}
Here, the lead-to-dot tunnel probability into a specific electronic dot state
with energy $E_q$ is proportional to the equilibrium population $\langle
c_{q,m}^\dagger c_{q,m}\rangle = f(E_q-\mu_R)$ with the source contact
characterized by the Fermi function $f(E-\mu_R) = [e^{(E-\mu_R)/\kT}+1]^{-1}$ at
temperature $T$ and chemical potential $\mu_R$.  For simplicity, we assume
within a wide-band approximation that the spectral density of the source contact
is energy independent and find the tunnel coupling $\Gamma_R(E) = 2\pi\sum_q
|V_q|^2 \delta(E-E_q) \equiv\Gamma_R$ between the right dot ond the source
contact.  The tunnel coupling between the left dot and the drain contact (left
lead) is defined accordingly.  Since the coupling term $V_q$ is sample dependent
and not a priory known (it can be tuned by gate voltages), we have determined
the effective dot-lead tunnel couplings, $\Gamma_L$ and $\Gamma_R$,
experimentally by independent dc-measurements without applying an rf-field. Note
that the decay rate used in the main article is $\Gamma_\text{out}=\Gamma_L$,
while the initialization rate $\Gamma_\text{in}$ combines $\Gamma_R$ with the
singlet-triplet couplings in the last term of the Hamiltonian in Eq.~\eqref{H_DQD}.

\subsubsection{System-bath Hamiltonian}\label{suppl:system-bath_hamiltonian}

A central aim of our study is to investigate
our two-electron charge qubit and its decoherence, caused by the coupling to a dissipating environment,
which is encoded in the LZSM pattern and its visibility. The details in
the experimentally observed fading of the LZSM pattern with increasing
temperature reveal two main environmental influences captured by $\lambda^\star$ and $\alpha_Z$: the first one is
an inhomogeneous broadening most likely caused by slow charge noise;
the second influence is the phonon bath \cite{Schinner2009,Granger2012a} which yields quantum dissipation and direct decoherence.  
Another possible decoherence source is the coupling to circuit noise which is important for typically impedance matched superconducting qubits \cite{Makhlin2001b}. In our case, however, we expect this external noise source to be of minor relevance owing to a strong impedance mismatch.
 
For the inhomogeneous broadening, we assume that it stems
from practically temperature independent slow fluctuations of the local
potential that remain constant during the typical dwell
time of an electron in the DQD. Therefore we can capture these fluctuations
by convoluting their amplitude distribution with $I(\bar\epsilon,A)$. 

For describing decoherence that stems from the interaction between
the DQD and bulk phonons, we employ a system-bath
approach in the spirit of the Caldeira-Leggett model for the
dissipative two-level system.  This means that we
couple the DQD to an ensemble of harmonic oscillators described by
the Hamiltonian $H_\text{bath} = \sum_\nu \hbar\omega_\nu
a_\nu^\dagger a_\nu$, where $a_\nu^\dagger$ and $a_\nu$ are the usual
bosonic creation and annihilation operators for a phonon of frequency
$\omega_\nu$.  The position operators of the bath oscillators couple
to the occupation difference between the left and the right dot, $Z =n_L-n_R$, according to
\begin{equation}
\label{bathcoupling}
H_\text{dot--bath} = \sum_\nu \lambda_\nu (a_\nu^\dagger+a_\nu) Z\,.
\end{equation}
This electron-phonon coupling Hamiltonian describes the interaction of the DQD
with environmental degrees of freedom. Its immediate effect is that fluctuations
in the environment detune the electronic states which, in turn, results in a
randomization of the relative phase in a superposition of states with distinct
charge distribution, in particular of the singlets representing our qubit. The
latter is therefore subject to decoherence. An important characteristic of a
dissipating bath is its spectral density $J(\omega) = \pi\sum_\nu|\lambda_\nu|^2
\delta(\omega-\omega_\nu)$.  As for the leads, we assume also for the phonon
bath a continuum limit and replace $J(\omega)$ by the Ohmic spectral density $J(\omega) = \pi\alpha_Z\omega/2$.  The
dimensionless electron-phonon coupling strength $\alpha_Z$ reflects the dissipation strength, which
together with the temperature parametrizes the decoherence due to the phonon
bath.  The Ohmic spectral density represents the natural choice 
which we have tested by performing additional numerical calculations using super-Ohmic spectral densities
$J(\omega)\propto\omega^{s+1}$ with $s>0$, which however failed to reproduce the experimentally observed fading of the LZSM pattern with increasing temperature.
A possible explanation for the good agreement of our
data with an Ohmic spectral phonon density is the quasi one-dimensional character of the electron-phonon interaction in our DQD sample: decoherence is mainly caused
by the one-dimensional subset of phonons with wavevector parallel to the line
connecting the two quantum dots.
For one-dimensional problems, the Ohmic spectral density of the
electron-phonon coupling is microscopically justified \cite{Weiss1999a}.

While $\alpha_Z$ couples to the diagonal of the Hamiltonian in Eq.~\eqref{H_DQD}, one may in addition consider the off-diagonal coupling term, namely the coupling between phonons and the interdot tunnel barrier. For this purpose, one introduces a further dot--bath Hamiltonian like that in
in Eq.~\eqref{bathcoupling} but with $Z$ replaced by $X =
\sum_{m=\uparrow,\downarrow} (c_{Lm}^\dagger c_{Rm} +c_{Rm}^\dagger c_{Lm})$ and
the coupling strength denoted by $\alpha_X$.  Unlike the bath-coupling via $Z$,
the bath now entails a fluctuating tunnel matrix element. Therefore, $\alpha_X$,
much more than $\alpha_Z$, drives transitions between the quantum dots by phonon
emission or absorption.  Analyzing this effect, we found a significant asymmetry of the current as function of
the detuning, which is in contrast to our experimental results. The quantitative comparison with our
measurements revealed that $\alpha_X$ is roughly two orders of
magnitude smaller than $\alpha_Z$.  In summary, $\alpha_X$ is of minor relevance
for the qubit decoherence and need not be taken into account.

\subsection{Charge qubit formed by two-electron singlet states}\label{suppl:charge_qubit}

The simplest implementation of a DQD charge qubit is a single electron that
tunnels between two dots. Nevertheless, here we consider the more complex case
of two electrons charging a DQD. For the sake of applications, the two-electron
state has the important advantage that it allows one to utilize both, charge and
spin degrees of freedoms in a single DQD. This opens up a number of interesting
possibilities, such as using either the singlet-singlet or one of the
singlet-triplet transitions to define a qubit or even to combine both by
subsequently sweeping through adjacent avoided crossings.  Furthermore, the
two-electron configuration constitutes the simplest possible many-body problem
which yields a theoretically more interesting system compared to a single
electron. Here, we focus on the two singlet states
\begin{align}
|\Sll\rangle ={}& c_{L\up}^\dagger c_{L\down}^\dagger |0\rangle ,
\\
|\Slr\rangle ={}& \frac{1}{\sqrt{2}} ( c_{L\up}^\dagger c_{R\down}^\dagger
+ c_{L\down}^\dagger c_{R\up}^\dagger ) |0\rangle ,
\end{align}
which span the Hilbert space of our qubit, where $|0\rangle$ is the uncharged
state of the DQD. In this singlet subspace, the double dot Hamiltonian defined in
Eq.~\eqref{H_DQD} reads
\begin{equation}
H_\text{qubit} =  \frac{\Delta}{2}\sigma_x + \frac{\epsilon}{2}\sigma_z
                 -\frac{\epsilon}{2}\mathbbm{1},
\end{equation}
(which is equivalent to Eq.~\eqref{hamiltonian}) with the unity
matrix $\mathbbm{1}$. The electron-phonon coupling operator defined in
Eq.~\eqref{bathcoupling} then contains $Z=\sigma_z$.
This leads us to the well-known spin-boson model with energy splitting
$E=\sqrt{\Delta^2+\epsilon^2}$ and dissipation strength $\alpha =
\alpha_Z$ defined in the usual way \cite{Leggett1987a}.

\subsubsection{Qubit decoherence}

The qubit reaches thermal equilibrium within the energy relaxation time $T_1$
while in the limit of weak dissipation, $\alpha_z \kT\ll E$, its pseudo-spin
performs coherent oscillations which decay exponentially within the
coherence time $T_2$ (assuming that the
electron-phonon coupling is the main decoherence mechanism).  From a
corresponding Bloch-Redfield master equation (see below), both decay
times can be determined \cite{Hanggi1990a, Makhlin2001b}:
\begin{align}
\label{T1}
T_1^{-1} ={}& \frac{\pi\alpha_Z}{\hbar} \frac{\Delta^2}{E}
              \coth\Big(\frac{E}{2\kT}\Big) ,
\\
\label{T2}
T_2^{-1} ={}& \frac{1}{2}T_1^{-1} + \frac{\pi\alpha_Z}{\hbar}
              \frac{2\kT\epsilon^2}{E^2} .
\end{align}
In the high-temperature limit, $\kT\gg E$, the decoherence rate is proportional
to the temperature: $T_2^{-1} =\pi\alpha_Z[1+\epsilon^2/E^2] \kT/\hbar$.
In the low-temperature limit, $\kT\ll E$, quantum fluctuations take over and
the coherence time becomes temperature independent, $T_2 =2T_1
=(2\hbar/\pi\alpha_Z)(E/\Delta^2)$.
For temperatures $T\gtrsim \hbar\Delta/k_B$, decoherence is
weakest near $\epsilon=0$, while for $|\epsilon|\gg\Delta$, the dephasing time
decays proportional to $1/\epsilon$. Thus, at these relatively high temperatures $\epsilon=0$ defines a sweet point
for quantum operations provided that the environment predominantly
couples via the occupation operator $Z=\sigma_z$, i.\,e., for
$\alpha_X\ll\alpha_Z$ as assumed in Eqs.~\eqref{T1} and \eqref{T2}.
In the opposite limit  $\alpha_X\gg\alpha_Z$ a likewise spin-boson model would
result in Eqs.~\eqref{T1} and \eqref{T2}, but with the parameters $\Delta$ and $\epsilon$
interchanged and $\alpha_Z$ replaced by $\alpha_X$.

Eq.~\eqref{T2} overestimates the coherence time of our specific charge qubit based on two-electron states in a DQD as it uses a two-level approximation which neglects spin flips and the triplet states.
In order to determine the qubit dephasing beyond the two-level approximation, we
follow the lines of Ref.\ \cite{Fonseca2004a} and employ the Bloch-Redfield
formalism (see \sect{Bloch-Redfield-Floquet_theory}).
In contrast to our transport calculations, we here disregard the dot-lead
couplings and consider a qubit in a closed DQD configuration. We aim at
obtaining the Liouville operator for the DQD coupled to the phonon bath and
further including spin flips. Decoherence is manifest in the transient decay of
off-diagonal density matrix elements. The coherence time, $T_2$, is
straightforwardly found by computing the eigenvalues of the Liouville operator.
To compute $T_2$ we, hence, evaluate the equation of motion of the total density
operator beyond the rotating-wave approximation (see below and Eq.~\eqref{BlochRedfield}), albeit using a time-independent Hamiltonian.
One finds a pair of eigenvalues with imaginary parts close
to $\pm E$ which correspond to coherent qubit oscillations.  Their real parts
are equal and can be identified as $T_2^{-1}$.

\begin{figure}
\begin{center}
\includegraphics[width=.95\columnwidth]{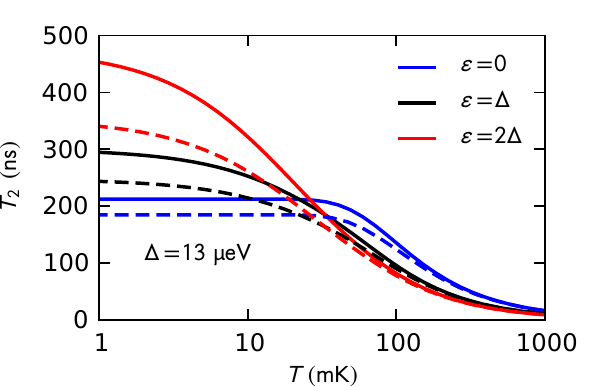}
\caption{\textbf{Temperature dependence of the qubit coherence.}
Decoherence time $T_2$ for our $\Sll$-$\Slr$ qubit for various values of the
detuning $\epsilon$.  The solid lines visualize Eq.~\eqref{T2} considering the
electron-phonon coupling in a two-level model, while the dashed lines are
computed numerically for the full DQD Hamiltonian which, in addition, takes the
triplet states and e.\,g.\ spin-flips into account.
}
\label{fig:t2analytic}
\end{center}
\end{figure}

In \fig{fig:t2analytic}{} we compare the coherence time $T_2$ as a function of
temperature calculated analytically with Eq.~\eqref{T2} (weakly dissipative spin-boson model, two-level approximation), on the one hand, and the numerical result of the complete problem, on the other hand.
This reveals that the two-level approximation overestimates $T_2$ by about 15\% since it cannot capture incoherent singlet-triplet transitions induced by
spin-flips and dephasing.
For the temperature range of of experiment, $T\gtrsim 20$\,mK, the
sweet point at $\epsilon=0$ is most favorable and predicts $T_2$ times up to
200\,ns which, however, become significantly smaller with increasing temperature
and bias.  In the sub milli-Kelvin regime, we observe the opposite.  There the
strongly biased situation corresponds to pure phase noise for which the
spin-boson model predicts $T_2\propto 1/k_BT$.  This leaves some room for
speculating about a coherence gain by further cooling.  With such extrapolation,
however, we leave the range in which our experiments justify the ohmic bath model.

\subsubsection{Advantage of a steady state experiment}

Previous measurements of the dephasing time \cite{Nowack2007a,Bluhm2011} relied
on an explicit time-trace of Ramsey fringes, where for each instance of time
a probability was reconstructed from a large number of destructive measurements
of a transient.
Such an averaging technique requires repeated identical
preparation.  Owing to the inhomogeneous broadening caused by slow noise, the
time-trace of the averaged probability oscillations obtained typically decays on a much
smaller time scale $T_2^*$, hence $T_2$ is not directly accessible.  Our data,
by contrast, are measured in the stationary state of a much simpler experiment.
In the resulting LZSM pattern, inhomegeneous broadening and decoherence are
manifest in separate ways. Crucially, as a consequence, $T_2^\star$ and $T_2$
can be distinguished in the analysis described in \sect{suppl:experiment} and the main article.

\subsection{Bloch-Redfield-Floquet theory}\label{Bloch-Redfield-Floquet_theory}

We aim at computing the time-averaged steady-state current through our strongly
driven DQD including an appreciable number of levels coupled to the various
environments, namely (i) the leads, (ii) slow charge noise, and (iii) the heat
bath. Moreover, dealing with two-electron states we have to include spin flips
which allow transitions between the triplet and singlet sub-spaces and resolve
spin blockade.  Our experimental results are consistent with the assumption that
all these incoherent processes occur on time scales much larger than those of
the coherent DQD dynamics, as is indicated by the following experimental
observations: (i) the coupling to the leads and the spin relaxation rate
ultimately determine the maximal current that we may observe.  The latter is
significantly smaller than the inter-dot tunnel frequency multiplied by the
elementary charge, i.\,e., $\Gamma_{L/R}\ll\Delta$.  (ii) Charge noise is rather
slow as compared to all these tunneling processes.  Therefore we can treat it
as disorder that is constant during the dwell time of an electron in the DQD.
In other words, it leads to an inhomogeneous broadening that does not affect the
decoherence dynamics of the electrons.  (iii) The appearance of an interference
pattern indicates that the inter-dot tunneling must be predominantly coherent
which excludes strong coupling to a heat bath.  This is confirmed by our finding
that the dimensionless dissipation strength is several orders of magnitude below
the crossover to the so-called incoherent tunneling regime \cite{Hanggi1990a}.
We can not a-priory exclude stronger coupling to a small number of
individual (tunneling) defects, but the fact that we did not find any memory
effects makes such a strong coupling scenario unlikely. 

\subsubsection{Floquet ansatz}

To cope with the complex problem outlined above we use a reduced density matrix
approach with the Floquet states of the driven system in the absence of the
environments as basis states.  These basis states already incorporate the
rf-modulation and, therefore, allow us to apply a rotating-wave approximation,
conveniently resulting in a time independent Liouville equation. This
perturbative approach is reliable under the assumption of only weakly coupled
environments. It has been applied in the past to both rf-driven dissipative
quantum systems \cite{Kohler1997a} and rf-driven quantum transport
\cite{Kohler2005a}.

Floquet theory exploits the fact that a periodically
time-dependent Schrödinger equation of the type
$i\hbar\partial|\psi\rangle/\partial t = H_\text{DQD}(t)|\psi\rangle$
possesses a complete set of solutions of the form $|\psi(t)\rangle =
e^{-i\omega t}|\phi(t)\rangle$, where $\hbar\omega$ is called
quasienergy.  The Floquet state $|\phi(t)\rangle =
|\phi(t+2\pi/\Omega)\rangle \equiv \sum_k e^{-ik\Omega
t}|\phi_{k}\rangle$ is characterized by
shareing the time-periodicity of the Hamiltonian.
Therefore, it can be represented as Fourier
series which,
importantly, allows an efficient numerical treatment.  In analogy
to quasimomenta in Bloch theory employed
for spatially periodic potentials, in Floquet theory the
quasienergies can be divided into Brillouin zones of equivalent
states.  Thus, it is sufficient to solve the eigenvalue problem within
one Brillouin zone, e.\,g.\ for $-\Omega/2 \leq \omega < \Omega/2$.
By inserting the Floquet ansatz into the Schrödinger equation, we
obtain the eigenvalue equation
\begin{equation}
\label{Floquet}
\Big(H_\text{DQD}(t) - i\hbar\frac{\partial}{\partial t}\Big)|\phi(t)\rangle
= \hbar\omega|\phi(t)\rangle
\end{equation}
from which we compute a complete set of Floquet states,
$\{|\phi_n\rangle\}$, and the corresponding quasienergies $\hbar\omega_n$\,.

\subsubsection{Bloch-Redfield theory}

An established technique for studying a quantum system in weak contact
with an environment is Bloch-Redfield theory.  It is based on a
treatment of the system-environment coupling operator $V$ within
second-order perturbation theory by which one finds for the total
density operator the equation of motion \cite{Blum1996a}
\begin{equation}
\label{BlochRedfield}
\dot R = -\frac{i}{\hbar} [H_\text{DQD}(t)+H_\text{env}, R]
-\frac{1}{\hbar^2}\int_0^\infty d\tau [V,[V(t-\tau,t), R]]\,.
\end{equation}
The particular form of the coupling operator in the interaction picture,
$V(t-\tau,t) = U_\text{DQD}^\dagger(t-\tau,t) V U_\text{DQD}(t-\tau,t)$, stems
from the explicit time-dependence of the central quantum system.  By tracing out
the environmental degrees of freedom, one obtains an equation of motion for the
reduced density operator of the system, $\rho$.  This step requires one to
specify the state of the environment.  Here, we assume that it is in the grand
canonical state $\rho_\text{env}^\text{(eq)}$ and that it is uncorrelated with
the system, such that the total density operator factorizes into a system and an
environment part, $R \simeq \rho\otimes\rho_\text{env}^\text{(eq)}$.  Under
this condition the decomposition into the Floquet basis provides a master
equation of the form $\dot\rho = -i(\omega_n-\omega_m)\rho_{mn} +
\sum_{n'm'}L_{nm,n'm'}\rho_{n'm'}$, where the Bloch-Redfield tensor
$L_{nm,n'm'}$ follows in a straightforward way from Eq.~\eqref{BlochRedfield}.  In the
last step we assume that all matrix elements $\rho_{nm}$ evolve much slower than
the rf-field, which allows us to replace the Bloch-Redfield tensor by its time
average. In this way we obtain a time-independent master equation describing the
time evolution of the population of the (time dependent) Floquet states.

We are exclusively interested in the steady state of this master equation, which
for weak dissipation eventually becomes diagonal in the Floquet basis.
Exploiting this knowledge, we set the off-diagonal matrix elements to zero and
arrive at a master equation of the form 
\begin{equation}
\label{master}
\dot\rho_{nn} = \sum_{n'}\Big(
W_{n\leftarrow n'} \rho_{n'n'} - W_{n'\leftarrow n} \rho_{nn} \Big)\,,
\end{equation}
where $\rho_{nn}$ are the populations of the Floquet states.
In the following we present the results for the transition rates
$W_{n\leftarrow n'}$ which are evaluated as sketched above.

\subsubsection{Coupling between the double quantum dot and the leads}

To calculate the tunnel coupling between the right lead and the right
quantum dot, we evaluate the coefficients of the master Eq.~\eqref{master} by
replacing $V$ in Eq.~\eqref{BlochRedfield} with the tunnel coupling between the right
dot and the right lead given by Eq.~\eqref{leadcoupling}.  After some algebra, we
arrive at the transitions rates
\begin{equation}
\label{Wleads}
\begin{split}
W_{n\leftarrow n'}^\text{leads} ={} &
\frac{\Gamma_R}{\hbar}\sum_k
\Big|\sum_{k'}\langle\phi_{n,k'+k}|c_R^\dagger|\phi_{n',k'}\rangle\Big|^2
\\ & \qquad \times
f(\omega_n-\omega_{n'}+k\Omega-\mu_R)
\\ &
+\frac{\Gamma_R}{\hbar}\sum_k\Big|\sum_{k'}\langle\phi_{n,k'+k}|c_R|\phi_{n',k'}\rangle\Big|^2
\\ & \qquad \times
(1-f(\omega_n-\omega_{n'}+k\Omega-\mu_R)) ,
\end{split}
\end{equation}
where the first term describes tunneling from the right lead to the right dot,
while the second term refers to the opposite process.  The corresponding
Liouvillian for coupling to the left lead is obtained in the same way with the
accordingly modified dot-lead Hamiltonian.
Owing to charge conservation, the time-averaged currents are the same at all
interfaces.  Here we evaluate it at the right dot-lead barrier and obtain it as
the difference between the terms that describe in Eq.~\eqref{Wleads} tunneling from the
lead to the dot and those describing the opposite process:
\begin{equation}
\begin{split}
J_{n\leftarrow n'} = {} &
-\frac{\Gamma_R}{\hbar}\sum_k
\Big|\sum_{k'}\langle\phi_{n,k'+k}|c_R^\dagger|\phi_{n',k'}\rangle\Big|^2
\\ & \qquad\times
f(\omega_n-\omega_{n'}+k\Omega-\mu_R)
\\ &
+\frac{\Gamma_R}{\hbar}\sum_k\Big|\sum_{k'}\langle\phi_{n,k'+k}|c_R|\phi_{n',k'}\rangle\Big|^2
\\ & \qquad\times
(1-f(\omega_n-\omega_{n'}+k\Omega-\mu_R)) ,
\end{split}
\end{equation}
where $J_{n\leftarrow n}=0$ due to vanishing matrix elements.  Note that this
expression can also be derived in a more formal way by introducing a
counting variable for the lead electrons before tracing them out and, thus, it
does not rely on any specific interpretation of the tunneling terms.

\subsubsection{Coupling between the qubit states and the heat bath}

A Liouvillian that describes the influence of the dissipating
environment on the DQD is derived by the same procedure, but using for
$V$ the electron-phonon Hamiltonian Eq.~\eqref{leadcoupling}.  We obtain
\begin{equation}
\begin{split}
W_{n\leftarrow n'}^\text{decoherence} = &
2\sum_k \Big|\sum_{k'}\langle\phi_{n,k'+k}|Z|\phi_{n',k'}\rangle\Big|^2
\\ & \times
N(\omega_n-\omega_n'+k\Omega),
\end{split}
\end{equation}
where $N(\omega) = J(\omega)n_\text{th}(\omega)$ with the bosonic
thermal occupation number $n_\text{th}(\omega) =
[e^{\hbar\omega/\kT}-1]^{-1}$.  In order to arrive at this convenient
form, we have defined $J(-\omega) = -J(\omega)$, while the Bose function was
extended by analytic continuation.

As already mentioned, a coherent tunnel process between the charge
configurations (1,1) and (2,0) requires that
the spin configuration of both states is equal, and in our case this includes
only the singlet states $\Slr$ and $\Sll$. The reason is, that the triplet
$\Tll$ is too high in energy owing to the large intradot exchange
interaction.  In turn, a direct transition of the triplet states with (1,1)
charge configuration to $\Sll$ is inhibited. As a consequence the transport
process comes to a standstill until a transition to the $\Slr$ singlet
occurs.  In our setup, this spin blockade is resolved by two mechanisms.  First,
the Zeeman field of the nanomagnet close to the DQD possesses an inhomogeneity
by which singlets and triplets mix. They form narrow avoided
crossings at which spin blockade is resolved and a current peak emerges.  This
effect is fully coherent and contained in our DQD Hamiltonian, Eq.~\eqref{H_DQD}.
Second, spin flips are induced by the hyperfine interaction with nuclear spins
in the GaAs matrix, which we treat as incoherent relaxation.
Therefore, thermal motion of the nuclear spins coupled via the hyperfine
interaction represent a further dissipative environment.  Our LZSM measurements
do not provide clear experimental hints on memory effects (besides for very
small modulation amplitudes $A$, where we find indications for dynamic nuclear spin polarization, see Sec.\ \ref{suppl:DNP}) or a significant temperature dependence of the spin flip rate
(within the probed temperature range). This allows us to avoid the theoretical
difficulties of choosing a particular microscopic model and to capture spin
flips by a Lindblad form with a rate $\gamma_\sigma$.  Therefore, we add to
the master equation the Liouvillean $L[\rho] = \frac{1}{2}\gamma_\sigma
\sum_{\ell,m=\up\down} (2S_{\ell,m}\rho S_{\ell,m}^\dagger-S_{\ell,m}^\dagger
S_{\ell,m}\rho -\rho S_{\ell,m}^\dagger S_{\ell,m})$ with the spin flip operator
$S_{\ell,m} = c_{\ell,\bar m}^\dagger c_{\ell,m}$
for an electron on dot $\ell$, where $m=\up,\down$ and $\bar m\neq m$.
Decomposition into Floquet states
followed by rotating-wave approximation yields the rate
\begin{equation}
W_{n\leftarrow n'}^\text{spinflip} =
\gamma_\sigma \sum_{m=\up\down}\sum_k
\Big|\sum_{k'}\langle\phi_{n,k'+k}|S_{L,m}|\phi_{n',k'}\rangle\Big|^2
\,.
\end{equation}

\begin{table*}
\begin{tabular}{lcrl}
\hline\hline
DQD parameter & value in $\mu$eV && determined by
\\\hline
bias voltage $V$			& 1000 && externally applied voltage
\\
intra dot Coulomb energy $U$ 		& $3500 \pm 350$ && charging diagram
\\
inter dot Coulomb interaction $U'$	& $820 \pm 80$ && charging diagram
\\
inter-dot tunnel coupling $\Delta$	& $13 \pm 1$ && spin funnel, see \fig{fig:couplings}{C}
\\
source-dot tunnel rate
$\Gamma_R$			& 0.1 && estimated from current, of minor relevance
\\
dot-drain tunnel rate $\Gamma_L$	& $2\times10^{-3} {}^\dagger$ && from current without spin blockade
\\
spin relaxation $\gamma_\sigma$	& $10^{-3}$ && from  current with spin blockade
\\
$\mathrm{T}_\pm-\Slr$ splitting		& $0.12^\ddag$ && Landau-Zener transition, see Fig.~\ref{fig:couplings}{A}
\\\hline\hline
external parameters & value in $\mu$eV && determined by
\\\hline
photon energy $\hbar\Omega$ & $6.2 / 11.4 / 18.7$ && at modulation frequency of $1.5 / 2.75 / 4.5$\,GHz
\\
mean Zeeman splitting $g\mB B$	& 4.2 && $g\mB B_\text{ext}$; $|g|=0.36$ and $B_\text{ext}=200\,$mT
\\
thermal energy $k_BT$		& 1.7 -- 40 && cryostat and electron temperature
\\\hline\hline
environmental influences & value && determined by
\\\hline
inhomogeneous broadening $\lambda^\star$	& $3.5\pm0.5\,\mu$eV && from broadening in $I(\bar\epsilon)$ peaks
\\
Caldeira-Leggett parameter
$\alpha_Z$			& $(1.5\pm0.5)\times10^{-4}$ && decay of lemon arcs, temperature dependence
\\
Caldeira-Leggett parameter
$\alpha_X$			& $<5\times10^{-6}$ && asymmetry of LZSM pattern
\\
\hline\hline
\end{tabular}
\caption{Parameters used for the numerical calculations. 
$^\dagger$ The data of \fig{figure2}{} and \fig{fig:frequencies}{} were
measured with a slightly smaller dot-drain rate and, accordingly, the numerical
data were computed with $\Gamma_R=1.2\times10^{-3}\,\mu$eV.
$^\ddag$ Note that the S-T splittings are reduced compared to $g\mB\Delta
B_{x,z}$ according to the weight of $\Sll$ in the singlet state. In our case
we have $g\mB\Delta B_x\simeq 0.2\,\mu$eV which reduces to the $\mathrm{T}_\pm-\Slr$
splitting of $0.12\,\mu$eV.
\label{table}}
\end{table*}
In our numerical approach to the steady-state average current, we
first compute the Floquet states which allows us to evaluate the
transition probabilities:
\begin{equation}
W_{n\leftarrow n'} = W_{n\leftarrow n'}^\text{leads} +
W_{n\leftarrow n'}^\text{decoherence} + W_{n\leftarrow
n'}^\text{spinflip}\,,
\end{equation}
so that we obtain a specific expression for the master equation
Eq.~\eqref{master}.  The steady-state solution of this master equation,
$\rho_{nn}^{(\infty)}$,
follows straigtforwardly from the condition $\dot\rho_{nn}^{(\infty)} = 0 = \sum_{n'} (W_{n\leftarrow
n'}\rho_{n'n'}^{(\infty)} -W_{n'\leftarrow
n}\rho_{nn}^{(\infty)})$ together with the trace condition
$\sum_n\rho_{nn}^{(\infty)}=1$.  Finally we arrive at the dc
current $I = \sum_{nn'} J_{n\leftarrow n'} \rho_{n'n'}^{(\infty)}$.


\section{System parameters}\label{suppl:table}

Table \ref{table} summarizes the most important parameters which characterize
our DQD, the two-electron qubit and its coupling to the environment and which we
used in our numerical calculations. For applying the scattering formula
Eq.~\eqref{LZSbasic}, we identified the initialization rate $\Gamma_\text{in}$ for the process $(1,0)\to\Slr$ with the spin relaxation rate $\gamma_\sigma$, while the decay $(2,0)\to(1,0)$ occurs at dot-draim rate so that $\Gamma_\text{out}=\Gamma_L$.


\end{document}